\documentclass[twocolumn,aps,prl,floatfix,superscriptaddress,nobibnotes,nofootinbib,altaffilsymbol]{revtex4-2}
\usepackage{amsmath,amssymb}
\usepackage{graphicx,float}
\usepackage{times,txfonts}
\usepackage{color}
\usepackage{multirow}
\usepackage{bbold}
\usepackage{color}
\usepackage{soul}
\usepackage{hyperref}
\usepackage{times,txfonts}
\usepackage{natbib}
\usepackage{nicefrac}
\usepackage{blindtext}
\usepackage[export]{adjustbox}
\usepackage{mathrsfs}
\usepackage{textcomp}
\usepackage{blkarray}
\usepackage{multirow}
\usepackage[normalem]{ulem}
\usepackage{tabu}

\newcommand{\perm}{\operatorname{perm}}

\newcommand{\bra}[1]{\left\langle #1 \right|}
\newcommand{\ket}[1]{\left| #1 \right\rangle}
\newcommand{\braket}[2]{\left\langle {#1{\left| \vphantom{#1 #2} \right.} #2} \right\rangle}

\renewcommand{\epsilon}{\varepsilon}

\def\VR{\kern-\arraycolsep\strut\vrule &\kern-\arraycolsep}
\def\vr{\kern-\arraycolsep & \kern-\arraycolsep}
\usepackage{sistyle}
\usepackage{cancel}

\usepackage{soul}
\definecolor{lightblue}{RGB}{185,210,248}
\sethlcolor{lightblue}

\begin{document}

%\title{GHZ-like and N00N State Generation via Multiphoton Interference of Vector Modes}
%\title{Multiphoton Entanglement Generation via Multi-Structured-Photon Interference}
\title{Multiphoton Entanglement via Structured-Photon Interference}

\author{Yifan Dong}
\affiliation{National Laboratory of Solid State Microstructures, School of Physics, and Collaborative Innovation Center of Advanced Microstructures, Nanjing University, Nanjing 210093, China }

\author{Yufeng Zhou}
\affiliation{National Laboratory of Solid State Microstructures, School of Physics, and Collaborative Innovation Center of Advanced Microstructures, Nanjing University, Nanjing 210093, China }

\author{Jiawei Zhao}
\affiliation{National Laboratory of Solid State Microstructures, School of Physics, and Collaborative Innovation Center of Advanced Microstructures, Nanjing University, Nanjing 210093, China }

\author{Tianyou Ying}
\affiliation{National Laboratory of Solid State Microstructures, School of Physics, and Collaborative Innovation Center of Advanced Microstructures, Nanjing University, Nanjing 210093, China }

\author{Ryan Hogan}
\email{ryan.hogan@nju.edu.cn}
\affiliation{National Laboratory of Solid State Microstructures, School of Physics, and Collaborative Innovation Center of Advanced Microstructures, Nanjing University, Nanjing 210093, China }

\author{Shining Zhu}
\affiliation{National Laboratory of Solid State Microstructures, School of Physics, and Collaborative Innovation Center of Advanced Microstructures, Nanjing University, Nanjing 210093, China }

\author{Xiaoqin Gao}
\email{xiaoqin.gao@nju.edu.cn}
\affiliation{National Laboratory of Solid State Microstructures, School of Physics, and Collaborative Innovation Center of Advanced Microstructures, Nanjing University, Nanjing 210093, China }

\begin{abstract}
As a strictly nonclassical phenomenon, multiphoton entanglement is widely used in fundamental quantum optics, quantum information processing, and quantum technologies. However, such entanglement generation remains challenging, as it typically requires complex optical networks with several optical elements and multiple spontaneous parametric down-conversion events in nonlinear crystals.
In this work, we generate and manipulate multiphoton entangled states through structured-photon interference in a multiport interferometer. 
We derive analytical expressions for multiphoton interference and establish the conditions for generating Greenberger–Horne–Zeilinger (GHZ)-like entangled states through coherent superpositions of complementary polarization configurations. Our scheme demonstrates favorable scalability, achieving theoretical success probabilities exceeding previous multiphoton entanglement-generation approaches by approximately three and five orders of magnitude for 10- and 12-photon inputs, respectively. As an illustrative example, we examine three structured photons interfering in a symmetric tritter, and numerically demonstrate how their spatially varying polarization structures govern the spatial distributions and relative weights of different GHZ-like components. By tailoring the input photons' internal structure, we can generate multiple entangled components in parallel and control them selectively without reconfiguring the interferometric network. Overall, these results demonstrate how the internal spatial structure of photons can be harnessed as an active resource in multiphoton quantum interference, enabling complex quantum state engineering.
\end{abstract}

\date{\today}
\maketitle

\section{Introduction}
{Multiphoton entanglement is a distinctive nonclassical phenomenon in quantum optics~\cite{Pan2012,erhard2020hd}. By establishing correlations among several photons, it provides a means of testing quantum mechanics beyond the two-particle systems that present increasingly complex correlations. Such states have played an important role in experimental tests of local realism~\cite{Pan2000} and have enabled important protocols in quantum communication, including quantum teleportation~\cite{Bouwmeester1997}, entanglement swapping~\cite{Pan1998EntanglementSwapping}, and quantum repeaters~\cite{Briegel1998}, as well as approaches to quantum computation based on linear optics~\cite{Knill2001,Kok2007}. Beyond applications, multiphoton entanglement also provides a means of exploring quantum correlations across the degrees of freedom (DOFs) of multiple particles. In photonic systems, this has attracted considerable interest, owing to the availability of multiple DOFs for encoding quantum information and the opportunities this provides for quantum communication and information processing~\cite{Malik2016,Huber2013,erhard2020hd}.} 

{Despite these advances, preparing multiphoton entangled states remains challenging and experimentally demanding. A common approach relies on multiple spontaneous parametric down-conversion (SPDC) events in carefully arranged optical networks~\cite{pan2001experimental,yang2022sequential,yao2012observation,wang2016experimental,zhong201812}. Such schemes often require multiple pair sources, cascaded optical processes, or complex interferometric configurations, where maintaining photon indistinguishability and relative phase control is inherently difficult. Moreover, such requirements become increasingly stringent as the photon number grows, often accompanied by a substantial reduction in generation probability. Although entanglement involving ten or more photons has been experimentally demonstrated~\cite{wang2016experimental,zhong201812}, as well as genuine three-photon interference~\cite{agne2017observation}, and high-dimensional photonic states beyond qubits~\cite{ErhardGHZ2018,hu2025observation,hu2020experimental}, more efficient and flexible approaches to multiphoton entanglement generation remain highly desirable. This motivates alternative strategies that exploit the intrinsic photon DOFs rather than relying solely on increasingly complex optical networks.}

{One such strategy exploits quantum interference between independently prepared photons. Hong–Ou–Mandel (HOM) is a fundamental example, producing photon bunching and coincidence-event suppression~\cite{Hong1987,Bouchard2021} through the coherent superposition of indistinguishable pathways. In multiport interferometers, multiphoton interference follows suit now considering several input and output modes, enabling engineering of collective quantum correlations through network configurations, photon distinguishability and output postselection~\cite{Tichy2015,Menssen2017,Shchesnovich2015,tillmann2015generalized}. These principles underpin studies ranging genuine three-photon interference ~\cite{campos2000three,mahrlein2015complete,agne2017observation}, to boson sampling~\cite{tillmann2013experimental,wang2017high,wang2018toward} and linear-optical quantum operations ~\cite{Knill2001,Kok2007}. While conventional approaches often seek to maximize photon indistinguishability, deliberately tailoring the internal states of individual photons offers an additional degree of control over collective multiphoton interference and the resulting quantum correlations.}

{Structured photons provide a particularly versatile platform for exploring this possibility. The availability of multiple photonic degrees of freedom enables the encoding and manipulation of complex internal quantum states~\cite{Malik2016,Huber2013,erhard2020hd}. In particular, coupling polarization and orbital angular momentum (OAM)~\cite{Allen1992,bliokh2015spin,erhard2018twisted} gives rise to vector-vortex modes with spatially varying polarization~\cite{Dambrosio2013,Cardano2012}, which can be readily engineered using spatially varying birefringent elements such as q-plates~\cite{Marrucci2006,Karimi2009}. Tailoring these spatial and polarization degrees of freedom enables control of photon internal states and their interference, as demonstrated in tunable two-photon interference~\cite{d2019tunable}, spatially structured quantum erasure~\cite{schiano2024engineering}, the characterization of entangled structured photons~\cite{Gao2024FullSpatial}, and the generation and characterization of complete polarization Bell bases~\cite{Gao2025BellBasis}. 
Extending these capabilities to multiphoton systems offers opportunities not only to engineer collective quantum correlations but also to introduce spatial programmability into multiphoton entanglement generation. However, how the internal structures of individual photons can be systematically exploited to control the generation, spatial distribution, and composition of multiphoton entangled states remains largely unexplored.}

{In this work, we introduce a general theoretical framework for generating and manipulating multiphoton entanglement through structured-photon interference in an $ N$-port interferometer. Using independently prepared photons with engineered internal states, we derive analytical expressions for multiphoton interference and establish the conditions for generating Greenberger–Horne–Zeilinger (GHZ)-like entangled states through coherent superpositions of complementary polarization configurations. Remarkably, our scheme achieves theoretical success probabilities exceeding those of previous multiphoton entanglement-generation approaches by approximately three and five orders of magnitude for 10- and 12-photon inputs, respectively~\cite{wang2016experimental,zhong201812}, highlighting its favorable scalability. Using three structured photons interfering in a symmetric tritter as an illustrative example, we numerically demonstrate how the coupling between polarization and spatial degrees of freedom governs the spatial distributions and relative amplitudes of different GHZ-like entangled components. By exploiting input photon internal structure, multiple entangled components can be generated in parallel and selectively controlled without reconfiguring the interferometric setup. These results establish structured-photon interference as a versatile approach to engineering multiphoton entanglement, with opportunities for scalable generation and spatially multiplexed quantum information processing~\cite{valencia2026large}.}

\section{Theoretical framework}
Consider $N$ single photons injected into an multiport linear
interferometer with $N$ ports. Input and output ports are labeled by $k= 1, 2, \ldots, N$ and $j = 1, 2, \ldots, N$, respectively. Before entering the interferometer, each input photon passes through a fully tuned q-plate coupling its OAM and polarization DOFs~\cite{Marrucci2006,Karimi2009,Gao2025BellBasis}.

Let us suppose that input photons are horizontally polarized, denoted by $\ket{H}$. Writing as a superposition of left- and right-circular polarization states and subsequently transforming them by q-plates~\cite{Marrucci2006,Gao2025BellBasis}, we arrive at 

\begin{equation}
\begin{aligned}
    &|\chi_k\rangle
    =
    \frac{i}{\sqrt{2}}
    \left(e^{-2iq_k\theta_k}\ket{R}
    +e^{2iq_k\theta_k}\ket{L}\right).
\end{aligned}
\label{eq:internal_state}
\end{equation}
representing the internal state of the $k$-th photon, with $q_k$ as topological charge and $\theta_k$ as azimuthal coordinate. Utilizing $\ket{\chi_k}$, we can define mutual overlap between two input photons, $k$ and $l$, using the Gram matrix, $G_{kl}=\langle \chi_k|\chi_l\rangle$, which characterizes photon distinguishability utilizing
internal DOFs~\cite{Tichy2015,Menssen2017}. In the ideal mode-orthogonality limit, photons prepared
with the same q-plate charge have identical internal states, while photons prepared with different q-plate charges are mutally orthogonal.

To define input photon states, $\ket{\psi_{\mathrm{in}}}$, we define the creation operator, $\hat a_{k,\chi_k}^{\dagger}$, for an input port $k$ with internal state $\ket{\chi_k}$ such that
\begin{equation}
    \ket{\psi_{\mathrm{in}}}
    =
    \prod_{k=1}^{N}
    \hat a_{k,\chi_k}^{\dagger}
    \ket{0}.
\label{eq:input_state}
\end{equation}
Input states are then passed through the multiport interferometer, which applies a unitary matrix $U_{kj}$, which we take as a symmetric Fourier interferometer~\cite{Zukowski1997}. With simple algebraic manipulation, we write the output state before postselection as
\begin{equation}
    |\psi_{\mathrm{out}}\rangle
    =
    \prod_{k=1}^{N}
    \left(
    \sum_{j=1}^{N}
    U_{kj}\hat b_{j,\chi_k}^{\dagger}
    \right)
    |0\rangle .
\label{eq:general_output_state}
\end{equation}
We then examine different postselection conditions that correspond to selecting different photon-number configurations and different internal states. 

\section{Generation of Multiphoton GHZ-like state}

\subsection{A. General framework for GHZ-like state generation}

To generate multiphoton GHZ-like states, we must consider the antibunching case. Projection onto this subspace gives
\begin{equation}
\ket{\widetilde{\psi}_{\mathrm{GHZ}}}
    =
    \sum_{\sigma\in S_N}
    \left(
    \prod_{j=1}^{N}
    U_{\sigma(j),j}
    \right)
    \prod_{j=1}^{N}
    \hat b_{j,\chi_{\sigma(j)}}^{\dagger}
    \ket{0} .
\label{eq:antibunching_state}
\end{equation}
Analyzing the polarization structure of the antibunching output, we can expand
Eq.~\eqref{eq:antibunching_state} in the local $H/V$ basis. For a
polarization string,
$\boldsymbol{\pi}=(\pi_1,\ldots,\pi_N)$, with
$\pi_j\in\{H,V\}$, the corresponding amplitude is
\begin{equation}
A_{\boldsymbol{\pi}}
=
\sum_{\sigma\in S_N}
\prod_{j=1}^{N}
U_{\sigma(j),j}
X_{\sigma(j),j}^{(\pi_j)},
\label{eq:polarization_string_amplitude}
\end{equation}
where
\begin{equation}
X_{kj}^{(H)}
=
F_{q_k}(r_j)\cos(2q_k\theta_j),
\quad
X_{kj}^{(V)}
=
F_{q_k}(r_j)\sin(2q_k\theta_j),
\label{eq:local_coefficients}
\end{equation}
and introduce the shorthand notation, $C_{kj}=\cos(2q_k\theta_j)$, and $S_{kj}=\sin(2q_k\theta_j)$.
Pairing each polarization string with its complementary string,
$\overline{\boldsymbol{\pi}}$,  defines GHZ-like components written as
\begin{equation}
|\mathrm{GHZ}_{\boldsymbol{\pi}}^{\pm}\rangle
=
\frac{1}{\sqrt{2}}
\left(
|\boldsymbol{\pi}\rangle
\pm
|\overline{\boldsymbol{\pi}}\rangle
\right),
\label{eq:ghz_like_basis}
\end{equation}
with unnormalized weights, $P_{\boldsymbol{\pi}}^{\pm}=\frac{1}{2}\left|A_{\boldsymbol{\pi}}\pm A_{\overline{\boldsymbol{\pi}}}\right|^2$. 

These weights vary spatially through interference between complementary polarization strings, governed by the q-plate charges and azimuthal detection coordinates, and normalized by $P_{\mathrm{GHZ}}$ for the postselected state. We now examine this probability of generating multiphoton GHZ-like entanglement, $P_{\mathrm{GHZ}}$, for cases of identical internal states and mutually orthogonal internal states.

\subsection{B. Generation probability and scalability}
Using Eq.~\eqref{eq:antibunching_state}, the probability of generating multiphoton GHZ-like entanglement can be written as 
\begin{equation}
    P_{\mathrm{GHZ}}
    =
    \langle
    \widetilde{\psi}_{\mathrm{GHZ}}
    |
    \widetilde{\psi}_{\mathrm{GHZ}}
    \rangle   =
    \sum_{\sigma,\tau\in S_N}
    \left[
    \prod_{j=1}^{N}
    U_{\sigma(j),j}
    U^{*}_{\tau(j),j}
    G_{\tau(j),\sigma(j)}
    \right],
\label{eq:antibunching_probability}
\end{equation}
which depends on the interference between paths and the
internal-state overlaps. Here, we see a coherent sum over pairs of multiphoton paths, where amplitudes and the Gram matrix describe path interference and interference strength between different paths, respectively.

Consider now photons with mutually orthogonal ($q_i\neq q_j$), or identical ($q_i = q_j$) internal states. For mutually orthogonal internal states, we have 
\begin{equation}
P_{\mathrm{GHZ}}^{q_i\neq q_j}=\frac{N!}{N^N},   
\end{equation}
which applies for all photon numbers, $N$. Interestingly, for identical internal states, we find a distinct dependence on photon number, where for odd $N$,
\begin{equation}
P_{\mathrm{GHZ}}^{q_i=q_j}=\left|\operatorname{perm}(U_N)\right|^2, 
\label{eq:antibunching_identical}
\end{equation}
and for even $N$, the probability vanishes. We also note that photon bunching is possible, with a probability of $P_{\mathrm{B}}^{q_i=q_j} = N!/N^{N-1}$, provided internal states are identical. Mutually orthogonal internal states give a much smaller bunching probability. More details on all cases can be found in the Supplementary Material.
% We then examine the dependence of photon number parity, shown in 

\begin{figure}[t!]
\centering
\includegraphics[width=1\columnwidth]{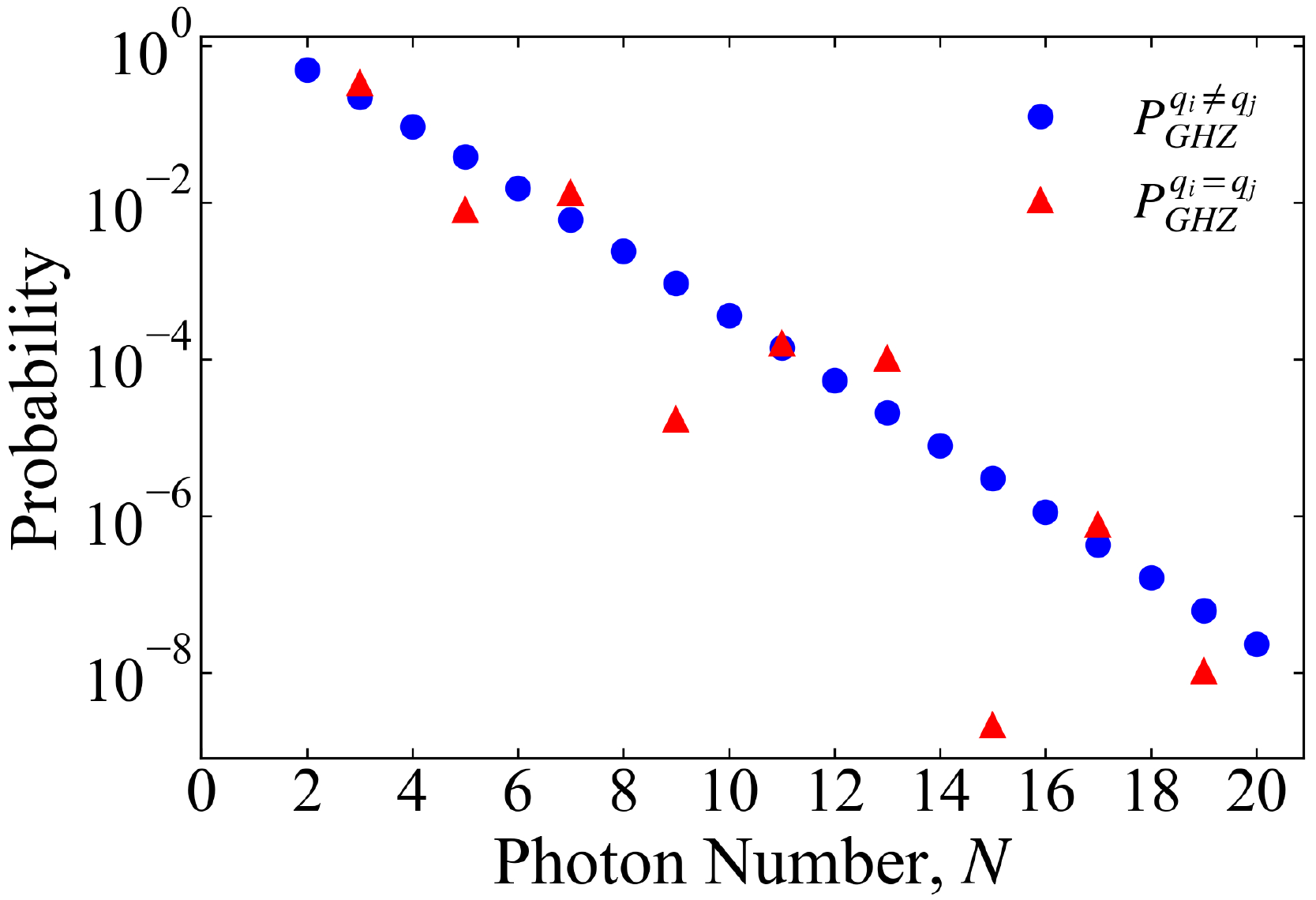}
\caption{\textbf{Photon-number dependence of GHZ-like entanglement generation probabilities.} For the antibunching case, postselection probability dependence on photon number, $N$, is shown for mutually orthogonal internal states ($q_i \neq q_j$, blue), corresponding to $P_{\mathrm{GHZ}}^{q_i\neq q_j}=N!/N^N$, and for identical internal
states ($q_i = q_j$, red), corresponding to 
$P_{\mathrm{GHZ}}^{q_i=q_j}=|\operatorname{perm}(U_N)|^2$.  For identical internal states, the probability vanishes for even $N$, resulting in zero-valued points that are not visible on the logarithmic scale. For mutually orthogonal internal states, probabilities are three and five orders of magnitude higher than the experimental approaches for multiphoton entanglement generation probabilities for 10- and 12-photon inputs, respectively~\cite{wang2016experimental,zhong201812}.}
\label{fig:postselection_probability_scaling}
\end{figure}

Examining photon number, Fig.~\ref{fig:postselection_probability_scaling} compares the dependence on scaling of multiphoton entanglement generation probabilities. Remarkably, for 10- and 12-photon inputs, our approach yields probabilities approximately three and five orders of magnitude higher, respectively, than those reported in previous experimental demonstrations~\cite{wang2016experimental,zhong201812}. These results highlight the significantly improved scaling of our approach with increasing photon number. With favorable photon-number scaling, we now turn to examine how photon internal structure can be exploited to control the composition of multiphoton entangled states. 

\begin{figure*}[t!]
    \centering \includegraphics[width=1\textwidth]{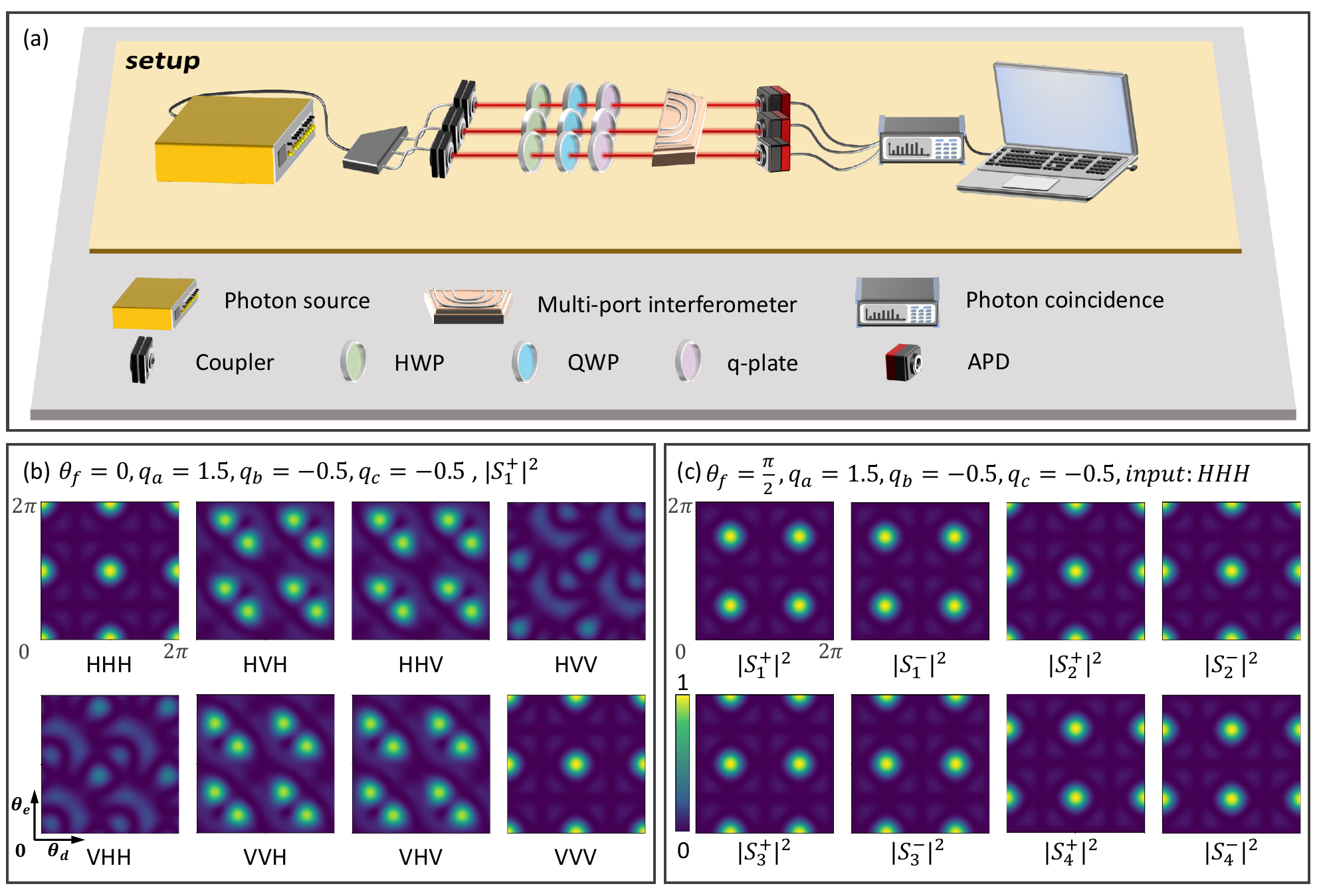}
\caption{
\textbf{Structured photon interference in a multiport interferometer.}
\textbf{(a)} Experimental scheme for generation and manipulation of multiphoton entanglement via structured-photon interference in a multiport interferometer. The setup comprises a photon source, couplers, polarization-control elements (half-wave plates (HWPs) and quarter-wave plates (QWPs)), $q$-plates, a multiport interferometer, avalanche photodiodes (APDs), and a photon-coincidence measurement system from which postselected events produce GHZ-like states. \textbf{(b)} Considering a three-photon example, we show simulated GHZ-like state probability spatial distributions for $|S_1^+|^2$ for different combinations of input photon polarizations (H or V), for $\theta_f=0$, $q_a=1.5$, $q_b=-0.5$, and $q_c=-0.5$.
\textbf{(c)} Simulated GHZ-like state probability spatial distribution for three input photons for all eight possible output GHZ-like states $|S_\mu^\pm|^2$ ($\mu=1,2,3,4$) is also shown, considering an HHH input state for $\theta_f=\pi/2$, with the same $q$-plate parameters as in (b).}
\label{fig:three_photon_setup}
\end{figure*}

%\section{Three-photon GHZ-like component analysis}
\section{Spatial engineering of Three-photon entanglement}
\label{sec:three_photon_ghz_component_analysis}
\subsection{A. Three-photon interference in a symmetric tritter}
As an example, we consider three structured photons interfering in a symmetric tritter, as shown in Fig.~\ref{fig:three_photon_setup} \textbf{(a)}. 
Using this configuration, we can investigate how spatially varying polarization structures influence the generation and spatial distribution of GHZ-like components. Preparation involves photons input into separate paths, which are then subsequently converted into vector modes by waveplates and q-plates. Vector photons are then injected into the symmetric tritter, with input and output ports labeled by $a,b,c$, and $d,e,f$, respectively.

We use the tritter unitary convention
% Postselecting events at the output ports allows us to analyze the resulting three-photon antibunching state and its GHZ-like polarization components.
\begin{equation}
    U^{(3)}
    =
    \frac{1}{\sqrt{3}}
    \begin{pmatrix}
        1 & 1 & 1 \\
        1 & \zeta^2 & \zeta \\
        1 & \zeta & \zeta^2
    \end{pmatrix},
    \qquad
    \zeta=e^{2\pi i/3},
    \label{eq:three_tritter_matrix}
\end{equation}
such that the general antibunching formula for three photons expanded in the local $H/V$ polarization basis gives
\begin{equation}
    |\widetilde{\psi}_{111}\rangle
    =
    \sum_{\boldsymbol{\pi}\in\{H,V\}^{3}}
    A_{\boldsymbol{\pi}}
    |\pi_d\pi_e\pi_f\rangle ,
    \label{eq:three_hv_expansion}
\end{equation}
with an associated output polarization string, $\boldsymbol{\pi}=(\pi_d,\pi_e,\pi_f)$.

\begin{figure*}
\centering
\includegraphics[width=1\textwidth]{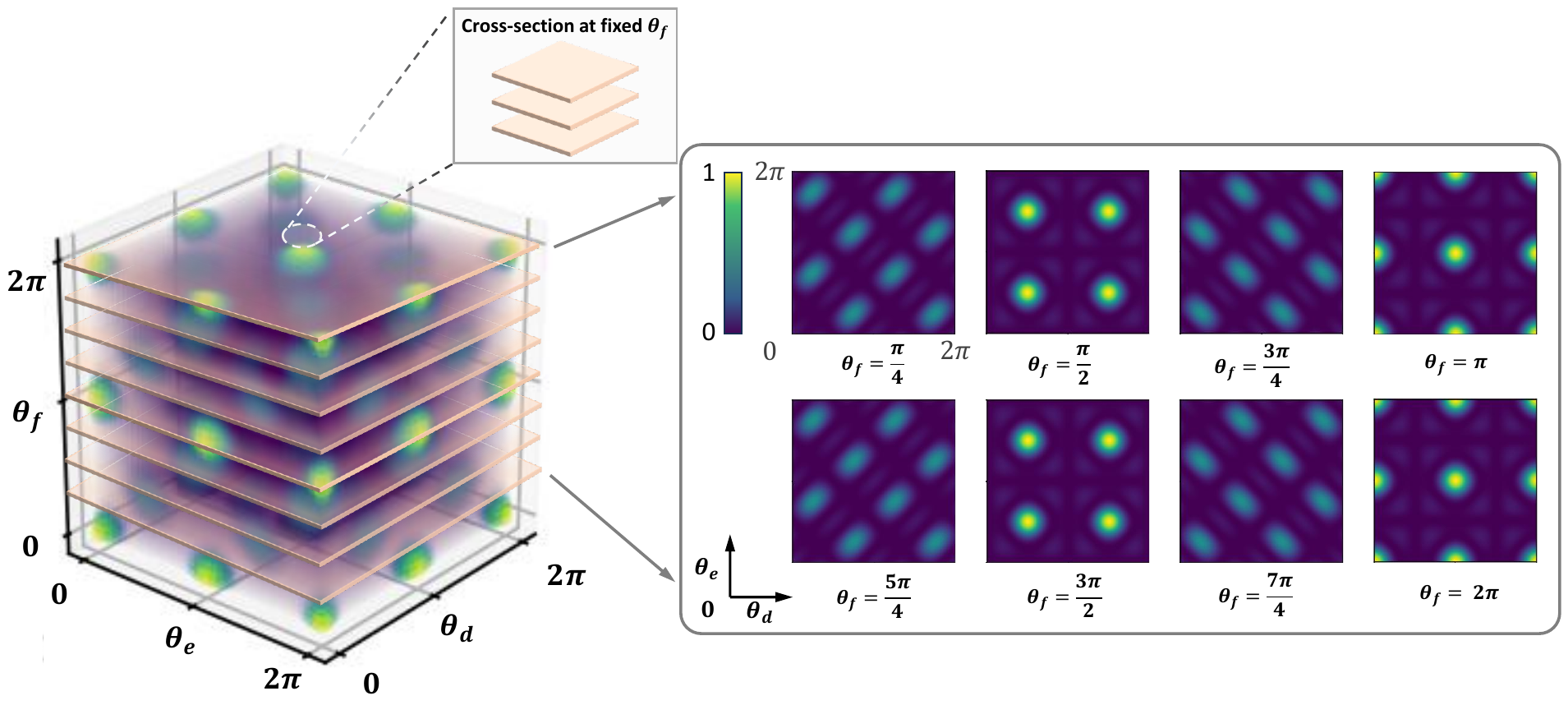}
\caption{
\textbf{Three-photon GHZ-like component $|S_1^+\rangle$ in three-dimensional azimuthal output space.}
The schematic on the left illustrates the spatial distribution of $|S_1^+|^2$ across the three azimuthal coordinates $\theta_d$, $\theta_e$, and $\theta_f$, divided into $\theta_f$ slices displayed on the right. The corresponding panels show two-dimensional cross-sections in the $(\theta_d,\theta_e)$ plane at fixed values of $\theta_f$, spanning one full $2\pi$ period. The evolution of the spatial patterns with $\theta_f$ reveals the periodic structure and symmetries of the three-photon output, with identical cross sections spaced by $\pi$ in $\theta_f$.}
\label{fig:three_s1p_theta_slices}
\end{figure*} 

  Suppose now that an input photon from port $k=x\in\{a,b,c\}$ with q-plate charge $q_x$ is
detected at output port $j\in\{d,e,f\}$.
% , we define
% \begin{equation}
%     C_{xj}
%     =
%     \cos(2q_x\theta_j),
%     \qquad
%     S_{xj}
%     =
%     \sin(2q_x\theta_j).
%     \label{eq:three_CS_definition}
% \end{equation}
If the input polarization in port $x$ is $p_x=\ket{H}$, local projection
factors are $X_{xj}^{H}=C_{xj}$, and $X_{xj}^{V}=S_{xj}$. If $p_x=\ket{V}$, we have $X_{xj}^{H}=S_{xj}$, and $X_{xj}^{V}=-C_{xj}$. Utilizing these expressions, each polarization string has a corresponding amplitude 
obtained by coherently summing over the $3!=6$ indistinguishable three-photon paths, and therefore written as
\begin{widetext}
\begin{equation}
    A_{\pi_d\pi_e\pi_f}
    =
    \frac{1}{3\sqrt{3}}
    \Big(
    \zeta
    X_{ad}^{\pi_d}
    X_{be}^{\pi_e}
    X_{cf}^{\pi_f}
    +
    \zeta^2
    X_{ad}^{\pi_d}
    X_{ce}^{\pi_e}
    X_{bf}^{\pi_f}+
    \zeta^2
    X_{bd}^{\pi_d}
    X_{ae}^{\pi_e}
    X_{cf}^{\pi_f}
    +
    \zeta
    X_{bd}^{\pi_d}
    X_{ce}^{\pi_e}
    X_{af}^{\pi_f}+
    \zeta
    X_{cd}^{\pi_d}
    X_{ae}^{\pi_e}
    X_{bf}^{\pi_f}
    +
    \zeta^2
    X_{cd}^{\pi_d}
    X_{be}^{\pi_e}
    X_{af}^{\pi_f}
    \Big).
    \label{eq:three_string_amplitude_compact}    
\end{equation}
\end{widetext}

% \begin{widetext}
% obtained by coherently summing over the $3!=6$ indistinguishable three-photon paths, and therefore written as
% \begin{equation}
% \begin{aligned}
%     A_{\pi_d\pi_e\pi_f}
%     =
%     &\frac{1}{3\sqrt{3}}
%     \Big(
%     \zeta
%     X_{ad}^{\pi_d}
%     X_{be}^{\pi_e}
%     X_{cf}^{\pi_f}
%     +
%     \zeta^2
%     X_{ad}^{\pi_d}
%     X_{ce}^{\pi_e}
%     X_{bf}^{\pi_f}+
%     \zeta^2
%     X_{bd}^{\pi_d}
%     X_{ae}^{\pi_e}
%     X_{cf}^{\pi_f}\\
%     & +
%     \zeta
%     X_{bd}^{\pi_d}
%     X_{ce}^{\pi_e}
%     X_{af}^{\pi_f}+
%     \zeta
%     X_{cd}^{\pi_d}
%     X_{ae}^{\pi_e}
%     X_{bf}^{\pi_f}
%     +
%     \zeta^2
%     X_{cd}^{\pi_d}
%     X_{be}^{\pi_e}
%     X_{af}^{\pi_f}
%     \Big).
%     \label{eq:three_string_amplitude_compact}
% \end{aligned}    
% \end{equation}
% \end{widetext}
The full expanded expressions for the eight polarization-string amplitudes are given in
the Supplementary Material.

With eight polarization strings in total, we group them into four complementary pairs and choose the three-photon GHZ-like basis as

\begin{align}
    |S_1^\pm\rangle
    &=
    \frac{1}{\sqrt{2}}
    \left(
    |HHH\rangle
    \pm
    |VVV\rangle
    \right),
    \;
    |S_2^\pm\rangle
    &=
    \frac{1}{\sqrt{2}}
    \left(
    |HHV\rangle
    \pm
    |VVH\rangle
    \right),
    \nonumber\\
    |S_3^\pm\rangle
    &=
    \frac{1}{\sqrt{2}}
    \left(
    |HVH\rangle
    \pm
    |VHV\rangle
    \right),
    \;
    |S_4^\pm\rangle
    &=
    \frac{1}{\sqrt{2}}
    \left(
    |HVV\rangle
    \pm
    |VHH\rangle
    \right),
    \label{eq:three_ghz_basis}
\end{align}
with corresponding coefficients as
\begin{align}
    S_1^\pm
    &=
    \frac{1}{\sqrt{2}}
    \left(
    A_{HHH}
    \pm
    A_{VVV}
    \right),
    \qquad
    S_2^\pm
    &=
    \frac{1}{\sqrt{2}}
    \left(
    A_{HHV}
    \pm
    A_{VVH}
    \right),
    \nonumber\\
    S_3^\pm
    &=
    \frac{1}{\sqrt{2}}
    \left(
    A_{HVH}
    \pm
    A_{VHV}
    \right),
    \qquad
    S_4^\pm
    &=
    \frac{1}{\sqrt{2}}
    \left(
    A_{HVV}
    \pm
    A_{VHH}
    \right).
    \label{eq:three_ghz_coefficients}
\end{align}
We characterize each GHZ-like component by the raw weight, $P_{S_\mu^\pm}=|S_\mu^\pm|^2$, with $\mu=1,2,3,4$. These quantities are not normalized at each spatial point but rather are unnormalized contributions to the corresponding GHZ-like component of the antibunching output state.

In the following subsection, we examine various dependencies of the spatial correlations by first beginning with the representative GHZ-like component in the three output azimuthal coordinates. 

% \subsection{Azimuthal-coordinate, polarization and topological charge dependencies}
% \label{subsec:three_angle_dependence}

\subsection{B. Spatially resolved GHZ-like entanglement}

We first consider
$|S_1^+\rangle$ as an example and calculate its probability, $P_{S_1^+}(\theta_d,\theta_e,\theta_f)=|S_1^+(\theta_d,\theta_e,\theta_f)|^2$ for different sets of input polarization, as shown in Fig.~\ref{fig:three_photon_setup} \textbf{(b)}. Taking $q_a=1.5$, $q_b=-0.5$, $q_c=-0.5$, and $\theta_f=\pi/2$,   we find that $\theta_e,\theta_d$-planes are spatially identical among polarization string pairs (for example, HHH and VVV, or HVV and VHH). Moreover, the choice of input polarization translates the positions of highest probability about the azimuthal plane. Considering only an HHH input, we examine the spatial probability distributions of all possible GHZ-like states, $P_{S_\mu^\pm}=|S_\mu^\pm|^2$, as shown in Fig.~\ref{fig:three_photon_setup} \textbf{(c)}. Each value of $\mu=1,2,3,4$ gives distinct spatial patterns, while the $\pm$-dependence for a given $\mu$ is identical, despite different polarization composition. Overall, we find a high degree of controllability of the engineering of the states and their associated symmetries. 

Another parameter that allows further engineering and control of the output states is depicted in Fig. ~\ref{fig:three_s1p_theta_slices}. Here, depicted in a three-dimensional coordinate space, $(\theta_d,\theta_e,\theta_f)$ alongside spatially distributed fixed-$\theta_f$ slices, we find that azimuthal dependence, $\theta_f$, does not simply rescale two-dimensional patterns in the $(\theta_d,\theta_e)$ plane, but changes the relative contributions of different paths based on coherent permutation of all paths from the tritter and sinusoidal factors
$\cos(2q_x\theta_j)$ and $\sin(2q_x\theta_j)$. The GHZ-like component maxima and minima are therefore redistributed in angular coordinate space. In addition to modulation by azimuthal coordinates, the topological charge further affects the response. 

\begin{figure}[t!]
\centering
\includegraphics[width=0.9\columnwidth]{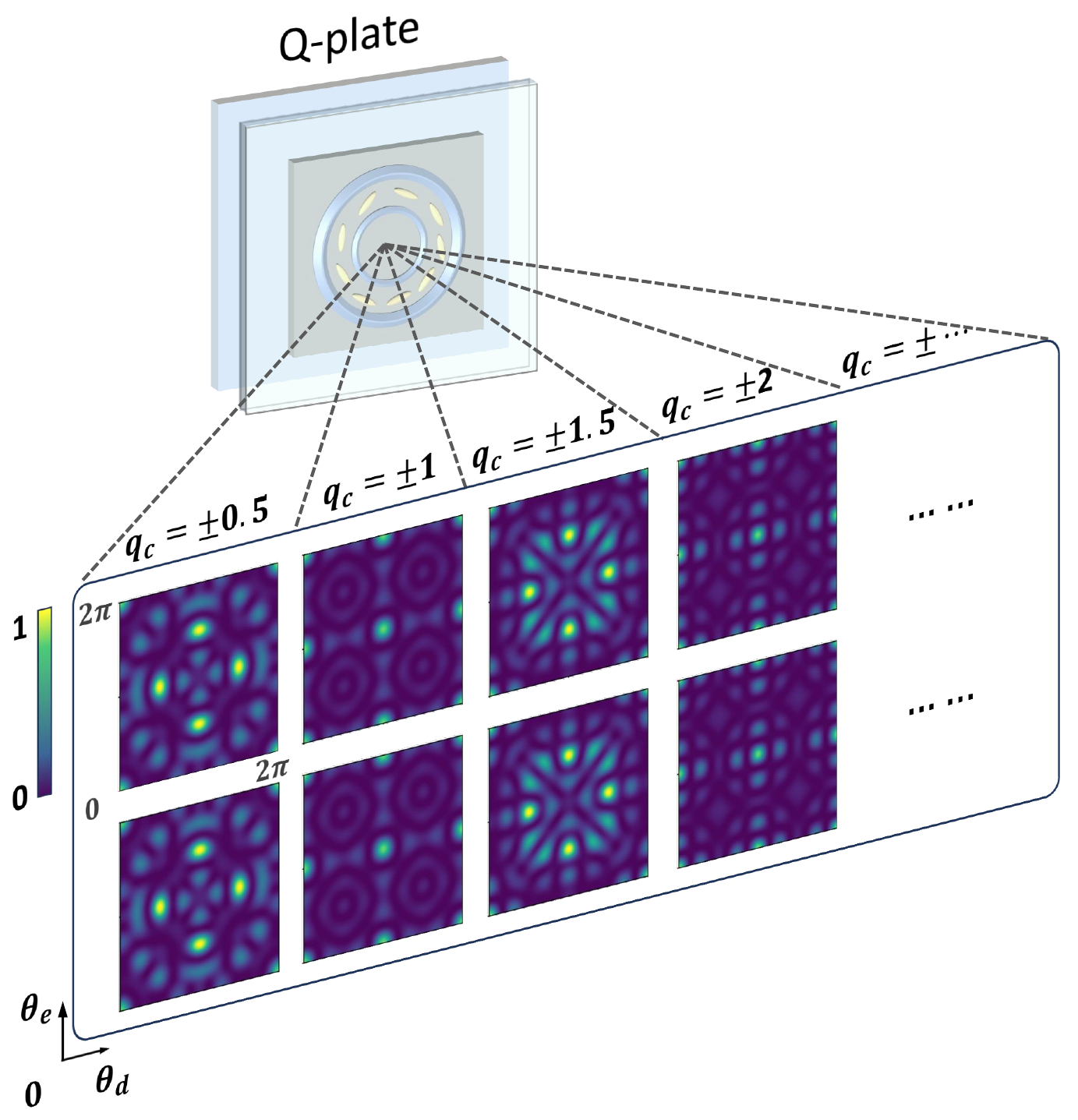}
\caption{\textbf{Topological charge dependence on three-photon GHZ-like $|S_1^+\rangle$ components.} Varying $q_c$ in steps of $\pm 0.5$ while holding $\theta_f=0$, $q_a=1$, and $q_b=2$, modifies the raw weight, $P_{S_1^+}(\theta_d,\theta_e)=|S_1^+(\theta_d,\theta_e)|^2$. This modification leads to modulation of multiplicity, spatial frequency, and overall spatial interference structure.  The top (bottom) row represents positive (negative) topological charge for $q_c$, highlighting that only the magnitude contributes to spatial distribution modification. }
\label{fig:three_s1p_qc_dependence}
\end{figure}

We now study a representative example of topological charge dependence by fixing $q_a=1$ and $q_b=2$, and varying $q_c$ in steps of $0.5$ to observe the effects on the GHZ-like components. Nonzero $q_c$ cases are shown for fixed-$\theta_f=0$ slices in Fig.~\ref{fig:three_s1p_qc_dependence}. We find that increasing $|q_c|$ introduces faster angular oscillations; however, since the postselected amplitude is a coherent sum over all three-photon permutation paths, changing $q_c$ modifies the full interference pattern rather than only a single output
coordinate. For $q_c=0$, azimuthal dependence associated with the third input photon disappears, and thus no longer possesses spatially varying polarization structure. 

Although not considered in this work, $q$-plate detuning can also play a role in engineering the output states. Detuning causes an incomplete conversion of Gaussian beams to vector vortex beams, leading to superposition input states. In this case, the additional free parameter significantly enhances the ability to engineer the GHZ-like components, providing greater tunability of the resulting multiphoton entanglement. Moreover, it has been well demonstrated that spatial patterns are easily modified using detuning~\cite{Gao2025BellBasis}, which inherently affects the underlying geometric phase~\cite{ying2026topological}. As such, detuning as a free parameter will be considered in future works to adjust spatial correlation patterns and the resulting entanglement into desired geometries. 

\subsection{C. Spatial Multiplexing and Control of GHZ-like Entanglement}

Beyond providing additional tunability, the spatial structure of the input photons offers another complementary DOF for engineering multiphoton entanglement. In particular, the ability to control the spatial distribution of individual GHZ-like components enables a form of point-by-point manipulation of entanglement across the detection plane, opening opportunities for parallel generation and spatial multiplexing. Different GHZ-like components can occupy distinct regions of the spatial detection domain, allowing their selective access through spatial detection alone. By choice of q-plate charges and input polarization states, along with azimuthal coordinate modulation, these spatial distributions can be controlled without modifying the interferometric network.

More broadly, this functionality arises from the interplay between the internal structures of individual photons and the coherent superposition of multiphoton interference pathways. Exploiting these intrinsic DOFs offers a potential route toward spatially multiplexed entanglement generation and programmable photonic quantum information processing.

\section{Conclusion}
In this work, we established a general theoretical framework for generating and manipulating multiphoton entanglement through structured-photon interference in a multiport interferometer. By exploiting the internal structures of individual photons, we derived analytical expressions for multiphoton interference and established the conditions for  GHZ-like entangled state generation through coherent superpositions of complementary polarization configurations. Importantly, our scheme exhibits a substantial improvement in success probability compared with previous approaches~\cite{wang2016experimental,zhong201812}, with the relative enhancement in success probability growing exponentially with photon number, highlighting its potential for scalable multiphoton entanglement generation.

Furthermore, using three structured photons interfering in a symmetric tritter, we numerically demonstrated how the coupling between polarization and spatial DOFs enables control over spatial distribution and relative weights of different GHZ-like entangled components. By modifying photon internal structure, these components can be selectively enhanced or suppressed without reconfiguring the interferometric network, offering a potential route toward spatially multiplexed multiphoton entanglement generation.

More broadly, our results establish individual photon internal structure as a controllable resource for engineering quantum states and multiphoton entanglement. Future investigations could extend this framework to larger multiphoton systems, examine its robustness against experimental imperfections such as partial distinguishability, photon loss, and q-plate detuning, and exploit additional photonic DOFs to synthesize more complex entangled states. These directions offer opportunities for advancing programmable photonic quantum information processing.

\section*{Acknowledgments}
The authors would like to acknowledge Armin Tavakoli, Robert Fickler, and Jessica Bavaresco for valuable scientific discussions.
This work was supported by the National Natural Science Foundation of China (Nos. 12574392, W2533027, U25A20194); the Natural Science Foundation of Jiangsu Province (Nos. BK20233001, BK20252118); and the Jiangsu Province Excellent Postdoctoral Fellowship Program (No. 2025ZB498). The authors also acknowledge support from the Jiangsu Physical Science Research Center.

%The authors also acknowledge support from the National Natural Science Foundation of China (Grant Nos. W2533027) and the Jiangsu Province Excellent Postdoctoral Fellowship Program (Grant No. 2025ZB498).

%\section*{Funding}
%The authors would like to acknowledge Robert Fickler and Armin Tavakoli for valuable scientific discussions. 

% \section*{Disclosures}
% The authors declare no conflicts of interest.

% \section*{Data Availability}
% Data underlying the results presented in this paper are not publicly available at this time but may be obtained from the authors upon reasonable request.

\bibliography{refs}

\begin{thebibliography}{44}%
\makeatletter
\providecommand \@ifxundefined [1]{%
 \@ifx{#1\undefined}
}%
\providecommand \@ifnum [1]{%
 \ifnum #1\expandafter \@firstoftwo
 \else \expandafter \@secondoftwo
 \fi
}%
\providecommand \@ifx [1]{%
 \ifx #1\expandafter \@firstoftwo
 \else \expandafter \@secondoftwo
 \fi
}%
\providecommand \natexlab [1]{#1}%
\providecommand \enquote  [1]{``#1''}%
\providecommand \bibnamefont  [1]{#1}%
\providecommand \bibfnamefont [1]{#1}%
\providecommand \citenamefont [1]{#1}%
\providecommand \href@noop [0]{\@secondoftwo}%
\providecommand \href [0]{\begingroup \@sanitize@url \@href}%
\providecommand \@href[1]{\@@startlink{#1}\@@href}%
\providecommand \@@href[1]{\endgroup#1\@@endlink}%
\providecommand \@sanitize@url [0]{\catcode `\\12\catcode `\$12\catcode
  `\&12\catcode `\#12\catcode `\^12\catcode `\_12\catcode `\%12\relax}%
\providecommand \@@startlink[1]{}%
\providecommand \@@endlink[0]{}%
\providecommand \url  [0]{\begingroup\@sanitize@url \@url }%
\providecommand \@url [1]{\endgroup\@href {#1}{\urlprefix }}%
\providecommand \urlprefix  [0]{URL }%
\providecommand \Eprint [0]{\href }%
\providecommand \doibase [0]{https://doi.org/}%
\providecommand \selectlanguage [0]{\@gobble}%
\providecommand \bibinfo  [0]{\@secondoftwo}%
\providecommand \bibfield  [0]{\@secondoftwo}%
\providecommand \translation [1]{[#1]}%
\providecommand \BibitemOpen [0]{}%
\providecommand \bibitemStop [0]{}%
\providecommand \bibitemNoStop [0]{.\EOS\space}%
\providecommand \EOS [0]{\spacefactor3000\relax}%
\providecommand \BibitemShut  [1]{\csname bibitem#1\endcsname}%
\let\auto@bib@innerbib\@empty
%</preamble>
\bibitem [{\citenamefont {Pan}\ \emph {et~al.}(2012)\citenamefont {Pan},
  \citenamefont {Chen}, \citenamefont {Lu}, \citenamefont {Weinfurter},
  \citenamefont {Zeilinger},\ and\ \citenamefont {Zukowski}}]{Pan2012}%
  \BibitemOpen
  \bibfield  {author} {\bibinfo {author} {\bibfnamefont {J.-W.}\ \bibnamefont
  {Pan}}, \bibinfo {author} {\bibfnamefont {Z.-B.}\ \bibnamefont {Chen}},
  \bibinfo {author} {\bibfnamefont {C.-Y.}\ \bibnamefont {Lu}}, \bibinfo
  {author} {\bibfnamefont {H.}~\bibnamefont {Weinfurter}}, \bibinfo {author}
  {\bibfnamefont {A.}~\bibnamefont {Zeilinger}},\ and\ \bibinfo {author}
  {\bibfnamefont {M.}~\bibnamefont {Zukowski}},\ }\bibfield  {title} {\bibinfo
  {title} {Multiphoton entanglement and interferometry},\ }\href
  {https://doi.org/10.1103/RevModPhys.84.777} {\bibfield  {journal} {\bibinfo
  {journal} {Reviews of Modern Physics}\ }\textbf {\bibinfo {volume} {84}},\
  \bibinfo {pages} {777} (\bibinfo {year} {2012})}\BibitemShut {NoStop}%
\bibitem [{\citenamefont {Erhard}\ \emph {et~al.}(2020)\citenamefont {Erhard},
  \citenamefont {Krenn},\ and\ \citenamefont {Zeilinger}}]{erhard2020hd}%
  \BibitemOpen
  \bibfield  {author} {\bibinfo {author} {\bibfnamefont {M.}~\bibnamefont
  {Erhard}}, \bibinfo {author} {\bibfnamefont {M.}~\bibnamefont {Krenn}},\ and\
  \bibinfo {author} {\bibfnamefont {A.}~\bibnamefont {Zeilinger}},\ }\bibfield
  {title} {\bibinfo {title} {Advances in high-dimensional quantum
  entanglement},\ }\href {https://doi.org/10.1038/s42254-020-0193-5} {\bibfield
   {journal} {\bibinfo  {journal} {Nature Reviews Physics}\ }\textbf {\bibinfo
  {volume} {2}},\ \bibinfo {pages} {365} (\bibinfo {year} {2020})}\BibitemShut
  {NoStop}%
\bibitem [{\citenamefont {Pan}\ \emph {et~al.}(2000)\citenamefont {Pan},
  \citenamefont {Bouwmeester}, \citenamefont {Daniell}, \citenamefont
  {Weinfurter},\ and\ \citenamefont {Zeilinger}}]{Pan2000}%
  \BibitemOpen
  \bibfield  {author} {\bibinfo {author} {\bibfnamefont {J.-W.}\ \bibnamefont
  {Pan}}, \bibinfo {author} {\bibfnamefont {D.}~\bibnamefont {Bouwmeester}},
  \bibinfo {author} {\bibfnamefont {M.}~\bibnamefont {Daniell}}, \bibinfo
  {author} {\bibfnamefont {H.}~\bibnamefont {Weinfurter}},\ and\ \bibinfo
  {author} {\bibfnamefont {A.}~\bibnamefont {Zeilinger}},\ }\bibfield  {title}
  {\bibinfo {title} {Experimental test of quantum nonlocality in three-photon
  greenberger-horne-zeilinger entanglement},\ }\href
  {https://doi.org/10.1038/35000514} {\bibfield  {journal} {\bibinfo  {journal}
  {Nature}\ }\textbf {\bibinfo {volume} {403}},\ \bibinfo {pages} {515}
  (\bibinfo {year} {2000})}\BibitemShut {NoStop}%
\bibitem [{\citenamefont {Bouwmeester}\ \emph {et~al.}(1997)\citenamefont
  {Bouwmeester}, \citenamefont {Pan}, \citenamefont {Mattle}, \citenamefont
  {Eibl}, \citenamefont {Weinfurter},\ and\ \citenamefont
  {Zeilinger}}]{Bouwmeester1997}%
  \BibitemOpen
  \bibfield  {author} {\bibinfo {author} {\bibfnamefont {D.}~\bibnamefont
  {Bouwmeester}}, \bibinfo {author} {\bibfnamefont {J.-W.}\ \bibnamefont
  {Pan}}, \bibinfo {author} {\bibfnamefont {K.}~\bibnamefont {Mattle}},
  \bibinfo {author} {\bibfnamefont {M.}~\bibnamefont {Eibl}}, \bibinfo {author}
  {\bibfnamefont {H.}~\bibnamefont {Weinfurter}},\ and\ \bibinfo {author}
  {\bibfnamefont {A.}~\bibnamefont {Zeilinger}},\ }\bibfield  {title} {\bibinfo
  {title} {Experimental quantum teleportation},\ }\href
  {https://doi.org/10.1038/37539} {\bibfield  {journal} {\bibinfo  {journal}
  {Nature}\ }\textbf {\bibinfo {volume} {390}},\ \bibinfo {pages} {575}
  (\bibinfo {year} {1997})}\BibitemShut {NoStop}%
\bibitem [{\citenamefont {Pan}\ \emph {et~al.}(1998)\citenamefont {Pan},
  \citenamefont {Bouwmeester}, \citenamefont {Weinfurter},\ and\ \citenamefont
  {Zeilinger}}]{Pan1998EntanglementSwapping}%
  \BibitemOpen
  \bibfield  {author} {\bibinfo {author} {\bibfnamefont {J.-W.}\ \bibnamefont
  {Pan}}, \bibinfo {author} {\bibfnamefont {D.}~\bibnamefont {Bouwmeester}},
  \bibinfo {author} {\bibfnamefont {H.}~\bibnamefont {Weinfurter}},\ and\
  \bibinfo {author} {\bibfnamefont {A.}~\bibnamefont {Zeilinger}},\ }\bibfield
  {title} {\bibinfo {title} {Experimental entanglement swapping: Entangling
  photons that never interacted},\ }\href
  {https://doi.org/10.1103/PhysRevLett.80.3891} {\bibfield  {journal} {\bibinfo
   {journal} {Physical Review Letters}\ }\textbf {\bibinfo {volume} {80}},\
  \bibinfo {pages} {3891} (\bibinfo {year} {1998})}\BibitemShut {NoStop}%
\bibitem [{\citenamefont {Briegel}\ \emph {et~al.}(1998)\citenamefont
  {Briegel}, \citenamefont {D{\"u}r}, \citenamefont {Cirac},\ and\
  \citenamefont {Zoller}}]{Briegel1998}%
  \BibitemOpen
  \bibfield  {author} {\bibinfo {author} {\bibfnamefont {H.-J.}\ \bibnamefont
  {Briegel}}, \bibinfo {author} {\bibfnamefont {W.}~\bibnamefont {D{\"u}r}},
  \bibinfo {author} {\bibfnamefont {J.~I.}\ \bibnamefont {Cirac}},\ and\
  \bibinfo {author} {\bibfnamefont {P.}~\bibnamefont {Zoller}},\ }\bibfield
  {title} {\bibinfo {title} {Quantum repeaters: The role of imperfect local
  operations in quantum communication},\ }\href
  {https://doi.org/10.1103/PhysRevLett.81.5932} {\bibfield  {journal} {\bibinfo
   {journal} {Physical Review Letters}\ }\textbf {\bibinfo {volume} {81}},\
  \bibinfo {pages} {5932} (\bibinfo {year} {1998})}\BibitemShut {NoStop}%
\bibitem [{\citenamefont {Knill}\ \emph {et~al.}(2001)\citenamefont {Knill},
  \citenamefont {Laflamme},\ and\ \citenamefont {Milburn}}]{Knill2001}%
  \BibitemOpen
  \bibfield  {author} {\bibinfo {author} {\bibfnamefont {E.}~\bibnamefont
  {Knill}}, \bibinfo {author} {\bibfnamefont {R.}~\bibnamefont {Laflamme}},\
  and\ \bibinfo {author} {\bibfnamefont {G.~J.}\ \bibnamefont {Milburn}},\
  }\bibfield  {title} {\bibinfo {title} {A scheme for efficient quantum
  computation with linear optics},\ }\href {https://doi.org/10.1038/35051009}
  {\bibfield  {journal} {\bibinfo  {journal} {Nature}\ }\textbf {\bibinfo
  {volume} {409}},\ \bibinfo {pages} {46} (\bibinfo {year} {2001})}\BibitemShut
  {NoStop}%
\bibitem [{\citenamefont {Kok}\ \emph {et~al.}(2007)\citenamefont {Kok},
  \citenamefont {Munro}, \citenamefont {Nemoto}, \citenamefont {Ralph},
  \citenamefont {Dowling},\ and\ \citenamefont {Milburn}}]{Kok2007}%
  \BibitemOpen
  \bibfield  {author} {\bibinfo {author} {\bibfnamefont {P.}~\bibnamefont
  {Kok}}, \bibinfo {author} {\bibfnamefont {W.~J.}\ \bibnamefont {Munro}},
  \bibinfo {author} {\bibfnamefont {K.}~\bibnamefont {Nemoto}}, \bibinfo
  {author} {\bibfnamefont {T.~C.}\ \bibnamefont {Ralph}}, \bibinfo {author}
  {\bibfnamefont {J.~P.}\ \bibnamefont {Dowling}},\ and\ \bibinfo {author}
  {\bibfnamefont {G.~J.}\ \bibnamefont {Milburn}},\ }\bibfield  {title}
  {\bibinfo {title} {Linear optical quantum computing with photonic qubits},\
  }\href {https://doi.org/10.1103/RevModPhys.79.135} {\bibfield  {journal}
  {\bibinfo  {journal} {Reviews of Modern Physics}\ }\textbf {\bibinfo {volume}
  {79}},\ \bibinfo {pages} {135} (\bibinfo {year} {2007})}\BibitemShut
  {NoStop}%
\bibitem [{\citenamefont {Malik}\ \emph {et~al.}(2016)\citenamefont {Malik},
  \citenamefont {Erhard}, \citenamefont {Huber}, \citenamefont {Krenn},
  \citenamefont {Fickler},\ and\ \citenamefont {Zeilinger}}]{Malik2016}%
  \BibitemOpen
  \bibfield  {author} {\bibinfo {author} {\bibfnamefont {M.}~\bibnamefont
  {Malik}}, \bibinfo {author} {\bibfnamefont {M.}~\bibnamefont {Erhard}},
  \bibinfo {author} {\bibfnamefont {M.}~\bibnamefont {Huber}}, \bibinfo
  {author} {\bibfnamefont {M.}~\bibnamefont {Krenn}}, \bibinfo {author}
  {\bibfnamefont {R.}~\bibnamefont {Fickler}},\ and\ \bibinfo {author}
  {\bibfnamefont {A.}~\bibnamefont {Zeilinger}},\ }\bibfield  {title} {\bibinfo
  {title} {Multi-photon entanglement in high dimensions},\ }\href
  {https://doi.org/10.1038/nphoton.2016.12} {\bibfield  {journal} {\bibinfo
  {journal} {Nature Photonics}\ }\textbf {\bibinfo {volume} {10}},\ \bibinfo
  {pages} {248} (\bibinfo {year} {2016})}\BibitemShut {NoStop}%
\bibitem [{\citenamefont {Huber}\ and\ \citenamefont
  {de~Vicente}(2013)}]{Huber2013}%
  \BibitemOpen
  \bibfield  {author} {\bibinfo {author} {\bibfnamefont {M.}~\bibnamefont
  {Huber}}\ and\ \bibinfo {author} {\bibfnamefont {J.~I.}\ \bibnamefont
  {de~Vicente}},\ }\bibfield  {title} {\bibinfo {title} {Structure of
  multidimensional entanglement in multipartite systems},\ }\href
  {https://doi.org/10.1103/PhysRevLett.110.030501} {\bibfield  {journal}
  {\bibinfo  {journal} {Physical Review Letters}\ }\textbf {\bibinfo {volume}
  {110}},\ \bibinfo {pages} {030501} (\bibinfo {year} {2013})}\BibitemShut
  {NoStop}%
\bibitem [{\citenamefont {Pan}\ \emph {et~al.}(2001)\citenamefont {Pan},
  \citenamefont {Daniell}, \citenamefont {Gasparoni}, \citenamefont {Weihs},\
  and\ \citenamefont {Zeilinger}}]{pan2001experimental}%
  \BibitemOpen
  \bibfield  {author} {\bibinfo {author} {\bibfnamefont {J.-W.}\ \bibnamefont
  {Pan}}, \bibinfo {author} {\bibfnamefont {M.}~\bibnamefont {Daniell}},
  \bibinfo {author} {\bibfnamefont {S.}~\bibnamefont {Gasparoni}}, \bibinfo
  {author} {\bibfnamefont {G.}~\bibnamefont {Weihs}},\ and\ \bibinfo {author}
  {\bibfnamefont {A.}~\bibnamefont {Zeilinger}},\ }\bibfield  {title} {\bibinfo
  {title} {Experimental demonstration of four-photon entanglement and
  high-fidelity teleportation},\ }\href
  {https://doi.org/10.1103/PhysRevLett.86.4435} {\bibfield  {journal} {\bibinfo
   {journal} {Physical Review Letters}\ }\textbf {\bibinfo {volume} {86}},\
  \bibinfo {pages} {4435} (\bibinfo {year} {2001})}\BibitemShut {NoStop}%
\bibitem [{\citenamefont {Yang}\ \emph {et~al.}(2022)\citenamefont {Yang},
  \citenamefont {Yu}, \citenamefont {Li}, \citenamefont {Jing}, \citenamefont
  {Bao},\ and\ \citenamefont {Pan}}]{yang2022sequential}%
  \BibitemOpen
  \bibfield  {author} {\bibinfo {author} {\bibfnamefont {C.-W.}\ \bibnamefont
  {Yang}}, \bibinfo {author} {\bibfnamefont {Y.}~\bibnamefont {Yu}}, \bibinfo
  {author} {\bibfnamefont {J.}~\bibnamefont {Li}}, \bibinfo {author}
  {\bibfnamefont {B.}~\bibnamefont {Jing}}, \bibinfo {author} {\bibfnamefont
  {X.-H.}\ \bibnamefont {Bao}},\ and\ \bibinfo {author} {\bibfnamefont {J.-W.}\
  \bibnamefont {Pan}},\ }\bibfield  {title} {\bibinfo {title} {Sequential
  generation of multiphoton entanglement with a rydberg superatom},\
  }\href@noop {} {\bibfield  {journal} {\bibinfo  {journal} {Nature Photonics}\
  }\textbf {\bibinfo {volume} {16}},\ \bibinfo {pages} {658} (\bibinfo {year}
  {2022})}\BibitemShut {NoStop}%
\bibitem [{\citenamefont {Yao}\ \emph {et~al.}(2012)\citenamefont {Yao},
  \citenamefont {Wang}, \citenamefont {Xu}, \citenamefont {Lu}, \citenamefont
  {Pan}, \citenamefont {Bao}, \citenamefont {Peng}, \citenamefont {Lu},
  \citenamefont {Chen},\ and\ \citenamefont {Pan}}]{yao2012observation}%
  \BibitemOpen
  \bibfield  {author} {\bibinfo {author} {\bibfnamefont {X.-C.}\ \bibnamefont
  {Yao}}, \bibinfo {author} {\bibfnamefont {T.-X.}\ \bibnamefont {Wang}},
  \bibinfo {author} {\bibfnamefont {P.}~\bibnamefont {Xu}}, \bibinfo {author}
  {\bibfnamefont {H.}~\bibnamefont {Lu}}, \bibinfo {author} {\bibfnamefont
  {G.-S.}\ \bibnamefont {Pan}}, \bibinfo {author} {\bibfnamefont {X.-H.}\
  \bibnamefont {Bao}}, \bibinfo {author} {\bibfnamefont {C.-Z.}\ \bibnamefont
  {Peng}}, \bibinfo {author} {\bibfnamefont {C.-Y.}\ \bibnamefont {Lu}},
  \bibinfo {author} {\bibfnamefont {Y.-A.}\ \bibnamefont {Chen}},\ and\
  \bibinfo {author} {\bibfnamefont {J.-W.}\ \bibnamefont {Pan}},\ }\bibfield
  {title} {\bibinfo {title} {Observation of eight-photon entanglement},\
  }\href@noop {} {\bibfield  {journal} {\bibinfo  {journal} {Nature photonics}\
  }\textbf {\bibinfo {volume} {6}},\ \bibinfo {pages} {225} (\bibinfo {year}
  {2012})}\BibitemShut {NoStop}%
\bibitem [{\citenamefont {Wang}\ \emph {et~al.}(2016)\citenamefont {Wang},
  \citenamefont {Chen}, \citenamefont {Li}, \citenamefont {Huang},
  \citenamefont {Liu}, \citenamefont {Chen}, \citenamefont {Luo}, \citenamefont
  {Su}, \citenamefont {Wu}, \citenamefont {Li} \emph
  {et~al.}}]{wang2016experimental}%
  \BibitemOpen
  \bibfield  {author} {\bibinfo {author} {\bibfnamefont {X.-L.}\ \bibnamefont
  {Wang}}, \bibinfo {author} {\bibfnamefont {L.-K.}\ \bibnamefont {Chen}},
  \bibinfo {author} {\bibfnamefont {W.}~\bibnamefont {Li}}, \bibinfo {author}
  {\bibfnamefont {H.-L.}\ \bibnamefont {Huang}}, \bibinfo {author}
  {\bibfnamefont {C.}~\bibnamefont {Liu}}, \bibinfo {author} {\bibfnamefont
  {C.}~\bibnamefont {Chen}}, \bibinfo {author} {\bibfnamefont {Y.-H.}\
  \bibnamefont {Luo}}, \bibinfo {author} {\bibfnamefont {Z.-E.}\ \bibnamefont
  {Su}}, \bibinfo {author} {\bibfnamefont {D.}~\bibnamefont {Wu}}, \bibinfo
  {author} {\bibfnamefont {Z.-D.}\ \bibnamefont {Li}}, \emph {et~al.},\
  }\bibfield  {title} {\bibinfo {title} {Experimental ten-photon
  entanglement},\ }\href {https://doi.org/10.1103/PhysRevLett.117.210502}
  {\bibfield  {journal} {\bibinfo  {journal} {Physical review letters}\
  }\textbf {\bibinfo {volume} {117}},\ \bibinfo {pages} {210502} (\bibinfo
  {year} {2016})}\BibitemShut {NoStop}%
\bibitem [{\citenamefont {Zhong}\ \emph {et~al.}(2018)\citenamefont {Zhong},
  \citenamefont {Li}, \citenamefont {Li}, \citenamefont {Peng}, \citenamefont
  {Su}, \citenamefont {Hu}, \citenamefont {He}, \citenamefont {Ding},
  \citenamefont {Zhang}, \citenamefont {Li} \emph {et~al.}}]{zhong201812}%
  \BibitemOpen
  \bibfield  {author} {\bibinfo {author} {\bibfnamefont {H.-S.}\ \bibnamefont
  {Zhong}}, \bibinfo {author} {\bibfnamefont {Y.}~\bibnamefont {Li}}, \bibinfo
  {author} {\bibfnamefont {W.}~\bibnamefont {Li}}, \bibinfo {author}
  {\bibfnamefont {L.-C.}\ \bibnamefont {Peng}}, \bibinfo {author}
  {\bibfnamefont {Z.-E.}\ \bibnamefont {Su}}, \bibinfo {author} {\bibfnamefont
  {Y.}~\bibnamefont {Hu}}, \bibinfo {author} {\bibfnamefont {Y.-M.}\
  \bibnamefont {He}}, \bibinfo {author} {\bibfnamefont {X.}~\bibnamefont
  {Ding}}, \bibinfo {author} {\bibfnamefont {W.}~\bibnamefont {Zhang}},
  \bibinfo {author} {\bibfnamefont {H.}~\bibnamefont {Li}}, \emph {et~al.},\
  }\bibfield  {title} {\bibinfo {title} {12-photon entanglement and scalable
  scattershot boson sampling with optimal entangled-photon pairs from
  parametric down-conversion},\ }\href
  {https://doi.org/10.1103/PhysRevLett.121.250505} {\bibfield  {journal}
  {\bibinfo  {journal} {Physical review letters}\ }\textbf {\bibinfo {volume}
  {121}},\ \bibinfo {pages} {250505} (\bibinfo {year} {2018})}\BibitemShut
  {NoStop}%
\bibitem [{\citenamefont {Agne}\ \emph {et~al.}(2017)\citenamefont {Agne},
  \citenamefont {Kauten}, \citenamefont {Jin}, \citenamefont {Meyer-Scott},
  \citenamefont {Salvail}, \citenamefont {Hamel}, \citenamefont {Resch},
  \citenamefont {Weihs},\ and\ \citenamefont
  {Jennewein}}]{agne2017observation}%
  \BibitemOpen
  \bibfield  {author} {\bibinfo {author} {\bibfnamefont {S.}~\bibnamefont
  {Agne}}, \bibinfo {author} {\bibfnamefont {T.}~\bibnamefont {Kauten}},
  \bibinfo {author} {\bibfnamefont {J.}~\bibnamefont {Jin}}, \bibinfo {author}
  {\bibfnamefont {E.}~\bibnamefont {Meyer-Scott}}, \bibinfo {author}
  {\bibfnamefont {J.~Z.}\ \bibnamefont {Salvail}}, \bibinfo {author}
  {\bibfnamefont {D.~R.}\ \bibnamefont {Hamel}}, \bibinfo {author}
  {\bibfnamefont {K.~J.}\ \bibnamefont {Resch}}, \bibinfo {author}
  {\bibfnamefont {G.}~\bibnamefont {Weihs}},\ and\ \bibinfo {author}
  {\bibfnamefont {T.}~\bibnamefont {Jennewein}},\ }\bibfield  {title} {\bibinfo
  {title} {Observation of genuine three-photon interference},\ }\href
  {https://doi.org/10.1103/PhysRevLett.118.153602} {\bibfield  {journal}
  {\bibinfo  {journal} {Physical review letters}\ }\textbf {\bibinfo {volume}
  {118}},\ \bibinfo {pages} {153602} (\bibinfo {year} {2017})}\BibitemShut
  {NoStop}%
\bibitem [{\citenamefont {Erhard}\ \emph
  {et~al.}(2018{\natexlab{a}})\citenamefont {Erhard}, \citenamefont {Malik},
  \citenamefont {Krenn} \emph {et~al.}}]{ErhardGHZ2018}%
  \BibitemOpen
  \bibfield  {author} {\bibinfo {author} {\bibfnamefont {M.}~\bibnamefont
  {Erhard}}, \bibinfo {author} {\bibfnamefont {M.}~\bibnamefont {Malik}},
  \bibinfo {author} {\bibfnamefont {M.}~\bibnamefont {Krenn}}, \emph {et~al.},\
  }\bibfield  {title} {\bibinfo {title} {Experimental
  greenberger--horne--zeilinger entanglement beyond qubits},\ }\href
  {https://doi.org/10.1038/s41566-018-0257-6} {\bibfield  {journal} {\bibinfo
  {journal} {Nature Photonics}\ }\textbf {\bibinfo {volume} {12}},\ \bibinfo
  {pages} {759} (\bibinfo {year} {2018}{\natexlab{a}})}\BibitemShut {NoStop}%
\bibitem [{\citenamefont {Hu}\ \emph {et~al.}(2025)\citenamefont {Hu},
  \citenamefont {Huang}, \citenamefont {d’Alessandro}, \citenamefont
  {Cobucci}, \citenamefont {Zhang}, \citenamefont {Guo}, \citenamefont {Huang},
  \citenamefont {Li}, \citenamefont {Guo}, \citenamefont {Gao} \emph
  {et~al.}}]{hu2025observation}%
  \BibitemOpen
  \bibfield  {author} {\bibinfo {author} {\bibfnamefont {X.-M.}\ \bibnamefont
  {Hu}}, \bibinfo {author} {\bibfnamefont {C.-X.}\ \bibnamefont {Huang}},
  \bibinfo {author} {\bibfnamefont {N.}~\bibnamefont {d’Alessandro}},
  \bibinfo {author} {\bibfnamefont {G.}~\bibnamefont {Cobucci}}, \bibinfo
  {author} {\bibfnamefont {C.}~\bibnamefont {Zhang}}, \bibinfo {author}
  {\bibfnamefont {Y.}~\bibnamefont {Guo}}, \bibinfo {author} {\bibfnamefont
  {Y.-F.}\ \bibnamefont {Huang}}, \bibinfo {author} {\bibfnamefont {C.-F.}\
  \bibnamefont {Li}}, \bibinfo {author} {\bibfnamefont {G.-C.}\ \bibnamefont
  {Guo}}, \bibinfo {author} {\bibfnamefont {X.}~\bibnamefont {Gao}}, \emph
  {et~al.},\ }\bibfield  {title} {\bibinfo {title} {Observation of genuine
  high-dimensional multi-partite non-locality in entangled photon states},\
  }\href {https://doi.org/10.1038/s41467-025-59717-y} {\bibfield  {journal}
  {\bibinfo  {journal} {Nature Communications}\ }\textbf {\bibinfo {volume}
  {16}},\ \bibinfo {pages} {5017} (\bibinfo {year} {2025})}\BibitemShut
  {NoStop}%
\bibitem [{\citenamefont {Hu}\ \emph {et~al.}(2020)\citenamefont {Hu},
  \citenamefont {Xing}, \citenamefont {Zhang}, \citenamefont {Liu},
  \citenamefont {Pivoluska}, \citenamefont {Huber}, \citenamefont {Huang},
  \citenamefont {Li},\ and\ \citenamefont {Guo}}]{hu2020experimental}%
  \BibitemOpen
  \bibfield  {author} {\bibinfo {author} {\bibfnamefont {X.}~\bibnamefont
  {Hu}}, \bibinfo {author} {\bibfnamefont {W.}~\bibnamefont {Xing}}, \bibinfo
  {author} {\bibfnamefont {C.}~\bibnamefont {Zhang}}, \bibinfo {author}
  {\bibfnamefont {B.}~\bibnamefont {Liu}}, \bibinfo {author} {\bibfnamefont
  {M.}~\bibnamefont {Pivoluska}}, \bibinfo {author} {\bibfnamefont
  {M.}~\bibnamefont {Huber}}, \bibinfo {author} {\bibfnamefont
  {Y.}~\bibnamefont {Huang}}, \bibinfo {author} {\bibfnamefont
  {C.}~\bibnamefont {Li}},\ and\ \bibinfo {author} {\bibfnamefont
  {G.}~\bibnamefont {Guo}},\ }\bibfield  {title} {\bibinfo {title}
  {Experimental creation of multi-photon high-dimensional layered quantum
  states. npj quan},\ }\href {https://doi.org/10.1038/s41534-020-00318-6}
  {\bibfield  {journal} {\bibinfo  {journal} {Inf}\ }\textbf {\bibinfo {volume}
  {6}},\ \bibinfo {pages} {88} (\bibinfo {year} {2020})}\BibitemShut {NoStop}%
\bibitem [{\citenamefont {Hong}\ \emph {et~al.}(1987)\citenamefont {Hong},
  \citenamefont {Ou},\ and\ \citenamefont {Mandel}}]{Hong1987}%
  \BibitemOpen
  \bibfield  {author} {\bibinfo {author} {\bibfnamefont {C.~K.}\ \bibnamefont
  {Hong}}, \bibinfo {author} {\bibfnamefont {Z.~Y.}\ \bibnamefont {Ou}},\ and\
  \bibinfo {author} {\bibfnamefont {L.}~\bibnamefont {Mandel}},\ }\bibfield
  {title} {\bibinfo {title} {Measurement of subpicosecond time intervals
  between two photons by interference},\ }\href
  {https://doi.org/10.1103/PhysRevLett.59.2044} {\bibfield  {journal} {\bibinfo
   {journal} {Physical Review Letters}\ }\textbf {\bibinfo {volume} {59}},\
  \bibinfo {pages} {2044} (\bibinfo {year} {1987})}\BibitemShut {NoStop}%
\bibitem [{\citenamefont {Bouchard}\ \emph {et~al.}(2021)\citenamefont
  {Bouchard}, \citenamefont {Sit}, \citenamefont {Zhang}, \citenamefont
  {Fickler}, \citenamefont {Miatto}, \citenamefont {Yao}, \citenamefont
  {Sciarrino}, \citenamefont {Boyd},\ and\ \citenamefont
  {Karimi}}]{Bouchard2021}%
  \BibitemOpen
  \bibfield  {author} {\bibinfo {author} {\bibfnamefont {F.}~\bibnamefont
  {Bouchard}}, \bibinfo {author} {\bibfnamefont {A.}~\bibnamefont {Sit}},
  \bibinfo {author} {\bibfnamefont {Y.}~\bibnamefont {Zhang}}, \bibinfo
  {author} {\bibfnamefont {R.}~\bibnamefont {Fickler}}, \bibinfo {author}
  {\bibfnamefont {F.~M.}\ \bibnamefont {Miatto}}, \bibinfo {author}
  {\bibfnamefont {E.~K.}\ \bibnamefont {Yao}}, \bibinfo {author} {\bibfnamefont
  {F.}~\bibnamefont {Sciarrino}}, \bibinfo {author} {\bibfnamefont {R.~W.}\
  \bibnamefont {Boyd}},\ and\ \bibinfo {author} {\bibfnamefont
  {E.}~\bibnamefont {Karimi}},\ }\bibfield  {title} {\bibinfo {title}
  {Two-photon interference: the hong--ou--mandel effect},\ }\href
  {https://doi.org/10.1088/1361-6633/abcd7a} {\bibfield  {journal} {\bibinfo
  {journal} {Reports on Progress in Physics}\ }\textbf {\bibinfo {volume}
  {84}},\ \bibinfo {pages} {012402} (\bibinfo {year} {2021})}\BibitemShut
  {NoStop}%
\bibitem [{\citenamefont {Tichy}(2015)}]{Tichy2015}%
  \BibitemOpen
  \bibfield  {author} {\bibinfo {author} {\bibfnamefont {M.~C.}\ \bibnamefont
  {Tichy}},\ }\bibfield  {title} {\bibinfo {title} {Sampling of partially
  distinguishable bosons and the relation to the multidimensional permanent},\
  }\href {https://doi.org/10.1103/PhysRevA.91.022316} {\bibfield  {journal}
  {\bibinfo  {journal} {Physical Review A}\ }\textbf {\bibinfo {volume} {91}},\
  \bibinfo {pages} {022316} (\bibinfo {year} {2015})}\BibitemShut {NoStop}%
\bibitem [{\citenamefont {Menssen}\ \emph {et~al.}(2017)\citenamefont
  {Menssen}, \citenamefont {Jones}, \citenamefont {Metcalf}, \citenamefont
  {Tichy}, \citenamefont {Barz}, \citenamefont {Kolthammer},\ and\
  \citenamefont {Walmsley}}]{Menssen2017}%
  \BibitemOpen
  \bibfield  {author} {\bibinfo {author} {\bibfnamefont {A.~J.}\ \bibnamefont
  {Menssen}}, \bibinfo {author} {\bibfnamefont {A.~E.}\ \bibnamefont {Jones}},
  \bibinfo {author} {\bibfnamefont {B.~J.}\ \bibnamefont {Metcalf}}, \bibinfo
  {author} {\bibfnamefont {M.~C.}\ \bibnamefont {Tichy}}, \bibinfo {author}
  {\bibfnamefont {S.}~\bibnamefont {Barz}}, \bibinfo {author} {\bibfnamefont
  {W.~S.}\ \bibnamefont {Kolthammer}},\ and\ \bibinfo {author} {\bibfnamefont
  {I.~A.}\ \bibnamefont {Walmsley}},\ }\bibfield  {title} {\bibinfo {title}
  {Distinguishability and many-particle interference},\ }\href
  {https://doi.org/10.1103/PhysRevLett.118.153603} {\bibfield  {journal}
  {\bibinfo  {journal} {Physical Review Letters}\ }\textbf {\bibinfo {volume}
  {118}},\ \bibinfo {pages} {153603} (\bibinfo {year} {2017})}\BibitemShut
  {NoStop}%
\bibitem [{\citenamefont {Shchesnovich}(2015)}]{Shchesnovich2015}%
  \BibitemOpen
  \bibfield  {author} {\bibinfo {author} {\bibfnamefont {V.~S.}\ \bibnamefont
  {Shchesnovich}},\ }\bibfield  {title} {\bibinfo {title} {Partial
  indistinguishability theory for multiphoton experiments in multiport
  devices},\ }\href {https://doi.org/10.1103/PhysRevA.91.013844} {\bibfield
  {journal} {\bibinfo  {journal} {Physical Review A}\ }\textbf {\bibinfo
  {volume} {91}},\ \bibinfo {pages} {013844} (\bibinfo {year}
  {2015})}\BibitemShut {NoStop}%
\bibitem [{\citenamefont {Tillmann}\ \emph {et~al.}(2015)\citenamefont
  {Tillmann}, \citenamefont {Tan}, \citenamefont {Stoeckl}, \citenamefont
  {Sanders}, \citenamefont {De~Guise}, \citenamefont {Heilmann}, \citenamefont
  {Nolte}, \citenamefont {Szameit},\ and\ \citenamefont
  {Walther}}]{tillmann2015generalized}%
  \BibitemOpen
  \bibfield  {author} {\bibinfo {author} {\bibfnamefont {M.}~\bibnamefont
  {Tillmann}}, \bibinfo {author} {\bibfnamefont {S.-H.}\ \bibnamefont {Tan}},
  \bibinfo {author} {\bibfnamefont {S.~E.}\ \bibnamefont {Stoeckl}}, \bibinfo
  {author} {\bibfnamefont {B.~C.}\ \bibnamefont {Sanders}}, \bibinfo {author}
  {\bibfnamefont {H.}~\bibnamefont {De~Guise}}, \bibinfo {author}
  {\bibfnamefont {R.}~\bibnamefont {Heilmann}}, \bibinfo {author}
  {\bibfnamefont {S.}~\bibnamefont {Nolte}}, \bibinfo {author} {\bibfnamefont
  {A.}~\bibnamefont {Szameit}},\ and\ \bibinfo {author} {\bibfnamefont
  {P.}~\bibnamefont {Walther}},\ }\bibfield  {title} {\bibinfo {title}
  {Generalized multiphoton quantum interference},\ }\href
  {https://doi.org/10.1103/PhysRevX.5.041015} {\bibfield  {journal} {\bibinfo
  {journal} {Physical Review X}\ }\textbf {\bibinfo {volume} {5}},\ \bibinfo
  {pages} {041015} (\bibinfo {year} {2015})}\BibitemShut {NoStop}%
\bibitem [{\citenamefont {Campos}(2000)}]{campos2000three}%
  \BibitemOpen
  \bibfield  {author} {\bibinfo {author} {\bibfnamefont {R.~A.}\ \bibnamefont
  {Campos}},\ }\bibfield  {title} {\bibinfo {title} {Three-photon
  hong-ou-mandel interference at a multiport mixer},\ }\href
  {https://doi.org/10.1103/PhysRevA.62.013809} {\bibfield  {journal} {\bibinfo
  {journal} {Physical Review A}\ }\textbf {\bibinfo {volume} {62}},\ \bibinfo
  {pages} {013809} (\bibinfo {year} {2000})}\BibitemShut {NoStop}%
\bibitem [{\citenamefont {M{\"a}hrlein}\ \emph {et~al.}(2015)\citenamefont
  {M{\"a}hrlein}, \citenamefont {Von~Zanthier},\ and\ \citenamefont
  {Agarwal}}]{mahrlein2015complete}%
  \BibitemOpen
  \bibfield  {author} {\bibinfo {author} {\bibfnamefont {S.}~\bibnamefont
  {M{\"a}hrlein}}, \bibinfo {author} {\bibfnamefont {J.}~\bibnamefont
  {Von~Zanthier}},\ and\ \bibinfo {author} {\bibfnamefont {G.~S.}\ \bibnamefont
  {Agarwal}},\ }\bibfield  {title} {\bibinfo {title} {Complete three photon
  hong-ou-mandel interference at a three port device},\ }\href
  {https://doi.org/10.1364/OE.23.015833} {\bibfield  {journal} {\bibinfo
  {journal} {Optics express}\ }\textbf {\bibinfo {volume} {23}},\ \bibinfo
  {pages} {15833} (\bibinfo {year} {2015})}\BibitemShut {NoStop}%
\bibitem [{\citenamefont {Tillmann}\ \emph {et~al.}(2013)\citenamefont
  {Tillmann}, \citenamefont {Daki{\'c}}, \citenamefont {Heilmann},
  \citenamefont {Nolte}, \citenamefont {Szameit},\ and\ \citenamefont
  {Walther}}]{tillmann2013experimental}%
  \BibitemOpen
  \bibfield  {author} {\bibinfo {author} {\bibfnamefont {M.}~\bibnamefont
  {Tillmann}}, \bibinfo {author} {\bibfnamefont {B.}~\bibnamefont {Daki{\'c}}},
  \bibinfo {author} {\bibfnamefont {R.}~\bibnamefont {Heilmann}}, \bibinfo
  {author} {\bibfnamefont {S.}~\bibnamefont {Nolte}}, \bibinfo {author}
  {\bibfnamefont {A.}~\bibnamefont {Szameit}},\ and\ \bibinfo {author}
  {\bibfnamefont {P.}~\bibnamefont {Walther}},\ }\bibfield  {title} {\bibinfo
  {title} {Experimental boson sampling},\ }\href
  {https://doi.org/10.1038/nphoton.2013.102} {\bibfield  {journal} {\bibinfo
  {journal} {Nature photonics}\ }\textbf {\bibinfo {volume} {7}},\ \bibinfo
  {pages} {540} (\bibinfo {year} {2013})}\BibitemShut {NoStop}%
\bibitem [{\citenamefont {Wang}\ \emph {et~al.}(2017)\citenamefont {Wang},
  \citenamefont {He}, \citenamefont {Li}, \citenamefont {Su}, \citenamefont
  {Li}, \citenamefont {Huang}, \citenamefont {Ding}, \citenamefont {Chen},
  \citenamefont {Liu}, \citenamefont {Qin} \emph {et~al.}}]{wang2017high}%
  \BibitemOpen
  \bibfield  {author} {\bibinfo {author} {\bibfnamefont {H.}~\bibnamefont
  {Wang}}, \bibinfo {author} {\bibfnamefont {Y.}~\bibnamefont {He}}, \bibinfo
  {author} {\bibfnamefont {Y.-H.}\ \bibnamefont {Li}}, \bibinfo {author}
  {\bibfnamefont {Z.-E.}\ \bibnamefont {Su}}, \bibinfo {author} {\bibfnamefont
  {B.}~\bibnamefont {Li}}, \bibinfo {author} {\bibfnamefont {H.-L.}\
  \bibnamefont {Huang}}, \bibinfo {author} {\bibfnamefont {X.}~\bibnamefont
  {Ding}}, \bibinfo {author} {\bibfnamefont {M.-C.}\ \bibnamefont {Chen}},
  \bibinfo {author} {\bibfnamefont {C.}~\bibnamefont {Liu}}, \bibinfo {author}
  {\bibfnamefont {J.}~\bibnamefont {Qin}}, \emph {et~al.},\ }\bibfield  {title}
  {\bibinfo {title} {High-efficiency multiphoton boson sampling},\ }\href
  {https://doi.org/10.1038/nphoton.2017.63} {\bibfield  {journal} {\bibinfo
  {journal} {Nature Photonics}\ }\textbf {\bibinfo {volume} {11}},\ \bibinfo
  {pages} {361} (\bibinfo {year} {2017})}\BibitemShut {NoStop}%
\bibitem [{\citenamefont {Wang}\ \emph {et~al.}(2018)\citenamefont {Wang},
  \citenamefont {Li}, \citenamefont {Jiang}, \citenamefont {He}, \citenamefont
  {Li}, \citenamefont {Ding}, \citenamefont {Chen}, \citenamefont {Qin},
  \citenamefont {Peng}, \citenamefont {Schneider} \emph
  {et~al.}}]{wang2018toward}%
  \BibitemOpen
  \bibfield  {author} {\bibinfo {author} {\bibfnamefont {H.}~\bibnamefont
  {Wang}}, \bibinfo {author} {\bibfnamefont {W.}~\bibnamefont {Li}}, \bibinfo
  {author} {\bibfnamefont {X.}~\bibnamefont {Jiang}}, \bibinfo {author}
  {\bibfnamefont {Y.-M.}\ \bibnamefont {He}}, \bibinfo {author} {\bibfnamefont
  {Y.-H.}\ \bibnamefont {Li}}, \bibinfo {author} {\bibfnamefont
  {X.}~\bibnamefont {Ding}}, \bibinfo {author} {\bibfnamefont {M.-C.}\
  \bibnamefont {Chen}}, \bibinfo {author} {\bibfnamefont {J.}~\bibnamefont
  {Qin}}, \bibinfo {author} {\bibfnamefont {C.-Z.}\ \bibnamefont {Peng}},
  \bibinfo {author} {\bibfnamefont {C.}~\bibnamefont {Schneider}}, \emph
  {et~al.},\ }\bibfield  {title} {\bibinfo {title} {Toward scalable boson
  sampling with photon loss},\ }\href
  {https://doi.org/10.1103/PhysRevLett.120.230502} {\bibfield  {journal}
  {\bibinfo  {journal} {Physical review letters}\ }\textbf {\bibinfo {volume}
  {120}},\ \bibinfo {pages} {230502} (\bibinfo {year} {2018})}\BibitemShut
  {NoStop}%
\bibitem [{\citenamefont {Allen}\ \emph {et~al.}(1992)\citenamefont {Allen},
  \citenamefont {Beijersbergen}, \citenamefont {Spreeuw},\ and\ \citenamefont
  {Woerdman}}]{Allen1992}%
  \BibitemOpen
  \bibfield  {author} {\bibinfo {author} {\bibfnamefont {L.}~\bibnamefont
  {Allen}}, \bibinfo {author} {\bibfnamefont {M.~W.}\ \bibnamefont
  {Beijersbergen}}, \bibinfo {author} {\bibfnamefont {R.~J.~C.}\ \bibnamefont
  {Spreeuw}},\ and\ \bibinfo {author} {\bibfnamefont {J.~P.}\ \bibnamefont
  {Woerdman}},\ }\bibfield  {title} {\bibinfo {title} {Orbital angular momentum
  of light and the transformation of laguerre-gaussian laser modes},\ }\href
  {https://doi.org/10.1103/PhysRevA.45.8185} {\bibfield  {journal} {\bibinfo
  {journal} {Physical Review A}\ }\textbf {\bibinfo {volume} {45}},\ \bibinfo
  {pages} {8185} (\bibinfo {year} {1992})}\BibitemShut {NoStop}%
\bibitem [{\citenamefont {Bliokh}\ \emph {et~al.}(2015)\citenamefont {Bliokh},
  \citenamefont {Rodr{\'\i}guez-Fortu{\~n}o}, \citenamefont {Nori},\ and\
  \citenamefont {Zayats}}]{bliokh2015spin}%
  \BibitemOpen
  \bibfield  {author} {\bibinfo {author} {\bibfnamefont {K.~Y.}\ \bibnamefont
  {Bliokh}}, \bibinfo {author} {\bibfnamefont {F.~J.}\ \bibnamefont
  {Rodr{\'\i}guez-Fortu{\~n}o}}, \bibinfo {author} {\bibfnamefont
  {F.}~\bibnamefont {Nori}},\ and\ \bibinfo {author} {\bibfnamefont {A.~V.}\
  \bibnamefont {Zayats}},\ }\bibfield  {title} {\bibinfo {title} {Spin--orbit
  interactions of light},\ }\href {https://doi.org/10.1038/nphoton.2015.201}
  {\bibfield  {journal} {\bibinfo  {journal} {Nature Photonics}\ }\textbf
  {\bibinfo {volume} {9}},\ \bibinfo {pages} {796} (\bibinfo {year}
  {2015})}\BibitemShut {NoStop}%
\bibitem [{\citenamefont {Erhard}\ \emph
  {et~al.}(2018{\natexlab{b}})\citenamefont {Erhard}, \citenamefont {Fickler},
  \citenamefont {Krenn},\ and\ \citenamefont {Zeilinger}}]{erhard2018twisted}%
  \BibitemOpen
  \bibfield  {author} {\bibinfo {author} {\bibfnamefont {M.}~\bibnamefont
  {Erhard}}, \bibinfo {author} {\bibfnamefont {R.}~\bibnamefont {Fickler}},
  \bibinfo {author} {\bibfnamefont {M.}~\bibnamefont {Krenn}},\ and\ \bibinfo
  {author} {\bibfnamefont {A.}~\bibnamefont {Zeilinger}},\ }\bibfield  {title}
  {\bibinfo {title} {Twisted photons: new quantum perspectives in high
  dimensions},\ }\href {https://doi.org/10.1038/lsa.2017.146} {\bibfield
  {journal} {\bibinfo  {journal} {Light: Science \& Applications}\ }\textbf
  {\bibinfo {volume} {7}},\ \bibinfo {pages} {17146} (\bibinfo {year}
  {2018}{\natexlab{b}})}\BibitemShut {NoStop}%
\bibitem [{\citenamefont {D'Ambrosio}\ \emph {et~al.}(2013)\citenamefont
  {D'Ambrosio}, \citenamefont {Spagnolo}, \citenamefont {Del~Re}, \citenamefont
  {Slussarenko}, \citenamefont {Li}, \citenamefont {Kwek}, \citenamefont
  {Marrucci}, \citenamefont {Walborn}, \citenamefont {Aolita},\ and\
  \citenamefont {Sciarrino}}]{Dambrosio2013}%
  \BibitemOpen
  \bibfield  {author} {\bibinfo {author} {\bibfnamefont {V.}~\bibnamefont
  {D'Ambrosio}}, \bibinfo {author} {\bibfnamefont {N.}~\bibnamefont
  {Spagnolo}}, \bibinfo {author} {\bibfnamefont {L.}~\bibnamefont {Del~Re}},
  \bibinfo {author} {\bibfnamefont {S.}~\bibnamefont {Slussarenko}}, \bibinfo
  {author} {\bibfnamefont {Y.}~\bibnamefont {Li}}, \bibinfo {author}
  {\bibfnamefont {L.~C.}\ \bibnamefont {Kwek}}, \bibinfo {author}
  {\bibfnamefont {L.}~\bibnamefont {Marrucci}}, \bibinfo {author}
  {\bibfnamefont {S.~P.}\ \bibnamefont {Walborn}}, \bibinfo {author}
  {\bibfnamefont {L.}~\bibnamefont {Aolita}},\ and\ \bibinfo {author}
  {\bibfnamefont {F.}~\bibnamefont {Sciarrino}},\ }\bibfield  {title} {\bibinfo
  {title} {Photonic polarization gears for ultra-sensitive angular
  measurements},\ }\href {https://doi.org/10.1038/ncomms3432} {\bibfield
  {journal} {\bibinfo  {journal} {Nature Communications}\ }\textbf {\bibinfo
  {volume} {4}},\ \bibinfo {pages} {2432} (\bibinfo {year} {2013})}\BibitemShut
  {NoStop}%
\bibitem [{\citenamefont {Cardano}\ \emph {et~al.}(2012)\citenamefont
  {Cardano}, \citenamefont {Karimi}, \citenamefont {Slussarenko}, \citenamefont
  {Marrucci}, \citenamefont {de~Lisio},\ and\ \citenamefont
  {Santamato}}]{Cardano2012}%
  \BibitemOpen
  \bibfield  {author} {\bibinfo {author} {\bibfnamefont {F.}~\bibnamefont
  {Cardano}}, \bibinfo {author} {\bibfnamefont {E.}~\bibnamefont {Karimi}},
  \bibinfo {author} {\bibfnamefont {S.}~\bibnamefont {Slussarenko}}, \bibinfo
  {author} {\bibfnamefont {L.}~\bibnamefont {Marrucci}}, \bibinfo {author}
  {\bibfnamefont {C.}~\bibnamefont {de~Lisio}},\ and\ \bibinfo {author}
  {\bibfnamefont {E.}~\bibnamefont {Santamato}},\ }\bibfield  {title} {\bibinfo
  {title} {Optical spin-to-orbital angular momentum conversion in inhomogeneous
  anisotropic media},\ }\href {https://doi.org/10.1103/PhysRevLett.108.073804}
  {\bibfield  {journal} {\bibinfo  {journal} {Physical Review Letters}\
  }\textbf {\bibinfo {volume} {108}},\ \bibinfo {pages} {073804} (\bibinfo
  {year} {2012})}\BibitemShut {NoStop}%
\bibitem [{\citenamefont {Marrucci}\ \emph {et~al.}(2006)\citenamefont
  {Marrucci}, \citenamefont {Manzo},\ and\ \citenamefont
  {Paparo}}]{Marrucci2006}%
  \BibitemOpen
  \bibfield  {author} {\bibinfo {author} {\bibfnamefont {L.}~\bibnamefont
  {Marrucci}}, \bibinfo {author} {\bibfnamefont {C.}~\bibnamefont {Manzo}},\
  and\ \bibinfo {author} {\bibfnamefont {D.}~\bibnamefont {Paparo}},\
  }\bibfield  {title} {\bibinfo {title} {Optical spin-to-orbital angular
  momentum conversion in inhomogeneous anisotropic media},\ }\href
  {https://doi.org/10.1103/PhysRevLett.96.163905} {\bibfield  {journal}
  {\bibinfo  {journal} {Physical Review Letters}\ }\textbf {\bibinfo {volume}
  {96}},\ \bibinfo {pages} {163905} (\bibinfo {year} {2006})}\BibitemShut
  {NoStop}%
\bibitem [{\citenamefont {Karimi}\ \emph {et~al.}(2009)\citenamefont {Karimi},
  \citenamefont {Piccirillo}, \citenamefont {Marrucci}, \citenamefont
  {Santamato},\ and\ \citenamefont {Nagali}}]{Karimi2009}%
  \BibitemOpen
  \bibfield  {author} {\bibinfo {author} {\bibfnamefont {E.}~\bibnamefont
  {Karimi}}, \bibinfo {author} {\bibfnamefont {B.}~\bibnamefont {Piccirillo}},
  \bibinfo {author} {\bibfnamefont {L.}~\bibnamefont {Marrucci}}, \bibinfo
  {author} {\bibfnamefont {E.}~\bibnamefont {Santamato}},\ and\ \bibinfo
  {author} {\bibfnamefont {E.}~\bibnamefont {Nagali}},\ }\bibfield  {title}
  {\bibinfo {title} {Light propagation in a birefringent plate with topological
  charge},\ }\href {https://doi.org/10.1364/OL.34.001225} {\bibfield  {journal}
  {\bibinfo  {journal} {Optics Letters}\ }\textbf {\bibinfo {volume} {34}},\
  \bibinfo {pages} {1225} (\bibinfo {year} {2009})}\BibitemShut {NoStop}%
\bibitem [{\citenamefont {D’Ambrosio}\ \emph {et~al.}(2019)\citenamefont
  {D’Ambrosio}, \citenamefont {Carvacho}, \citenamefont {Agresti},
  \citenamefont {Marrucci},\ and\ \citenamefont {Sciarrino}}]{d2019tunable}%
  \BibitemOpen
  \bibfield  {author} {\bibinfo {author} {\bibfnamefont {V.}~\bibnamefont
  {D’Ambrosio}}, \bibinfo {author} {\bibfnamefont {G.}~\bibnamefont
  {Carvacho}}, \bibinfo {author} {\bibfnamefont {I.}~\bibnamefont {Agresti}},
  \bibinfo {author} {\bibfnamefont {L.}~\bibnamefont {Marrucci}},\ and\
  \bibinfo {author} {\bibfnamefont {F.}~\bibnamefont {Sciarrino}},\ }\bibfield
  {title} {\bibinfo {title} {Tunable two-photon quantum interference of
  structured light},\ }\href {https://doi.org/10.1103/PhysRevLett.122.013601}
  {\bibfield  {journal} {\bibinfo  {journal} {Physical Review Letters}\
  }\textbf {\bibinfo {volume} {122}},\ \bibinfo {pages} {013601} (\bibinfo
  {year} {2019})}\BibitemShut {NoStop}%
\bibitem [{\citenamefont {Schiano}\ \emph {et~al.}(2024)\citenamefont
  {Schiano}, \citenamefont {Sephton}, \citenamefont {Aiello}, \citenamefont
  {Graffitti}, \citenamefont {Lal}, \citenamefont {Chiuri}, \citenamefont
  {Santoro}, \citenamefont {Amato}, \citenamefont {Marrucci}, \citenamefont
  {De~Lisio} \emph {et~al.}}]{schiano2024engineering}%
  \BibitemOpen
  \bibfield  {author} {\bibinfo {author} {\bibfnamefont {C.}~\bibnamefont
  {Schiano}}, \bibinfo {author} {\bibfnamefont {B.}~\bibnamefont {Sephton}},
  \bibinfo {author} {\bibfnamefont {R.}~\bibnamefont {Aiello}}, \bibinfo
  {author} {\bibfnamefont {F.}~\bibnamefont {Graffitti}}, \bibinfo {author}
  {\bibfnamefont {N.}~\bibnamefont {Lal}}, \bibinfo {author} {\bibfnamefont
  {A.}~\bibnamefont {Chiuri}}, \bibinfo {author} {\bibfnamefont
  {S.}~\bibnamefont {Santoro}}, \bibinfo {author} {\bibfnamefont {L.~S.}\
  \bibnamefont {Amato}}, \bibinfo {author} {\bibfnamefont {L.}~\bibnamefont
  {Marrucci}}, \bibinfo {author} {\bibfnamefont {C.}~\bibnamefont {De~Lisio}},
  \emph {et~al.},\ }\bibfield  {title} {\bibinfo {title} {Engineering quantum
  states from a spatially structured quantum eraser},\ }\href
  {https://doi.org/10.1126/sciadv.adm9278} {\bibfield  {journal} {\bibinfo
  {journal} {Science Advances}\ }\textbf {\bibinfo {volume} {10}},\ \bibinfo
  {pages} {eadm9278} (\bibinfo {year} {2024})}\BibitemShut {NoStop}%
\bibitem [{\citenamefont {Gao}\ \emph {et~al.}(2024)\citenamefont {Gao},
  \citenamefont {Zhang}, \citenamefont {D'Errico}, \citenamefont {Sit},
  \citenamefont {Heshami},\ and\ \citenamefont {Karimi}}]{Gao2024FullSpatial}%
  \BibitemOpen
  \bibfield  {author} {\bibinfo {author} {\bibfnamefont {X.}~\bibnamefont
  {Gao}}, \bibinfo {author} {\bibfnamefont {Y.}~\bibnamefont {Zhang}}, \bibinfo
  {author} {\bibfnamefont {A.}~\bibnamefont {D'Errico}}, \bibinfo {author}
  {\bibfnamefont {A.}~\bibnamefont {Sit}}, \bibinfo {author} {\bibfnamefont
  {K.}~\bibnamefont {Heshami}},\ and\ \bibinfo {author} {\bibfnamefont
  {E.}~\bibnamefont {Karimi}},\ }\bibfield  {title} {\bibinfo {title} {Full
  spatial characterization of entangled structured photons},\ }\href
  {https://doi.org/10.1103/PhysRevLett.132.063802} {\bibfield  {journal}
  {\bibinfo  {journal} {Physical Review Letters}\ }\textbf {\bibinfo {volume}
  {132}},\ \bibinfo {pages} {063802} (\bibinfo {year} {2024})}\BibitemShut
  {NoStop}%
\bibitem [{\citenamefont {Gao}\ \emph {et~al.}(2025)\citenamefont {Gao},
  \citenamefont {Paneru}, \citenamefont {Di~Colandrea}, \citenamefont {Zhang},\
  and\ \citenamefont {Karimi}}]{Gao2025BellBasis}%
  \BibitemOpen
  \bibfield  {author} {\bibinfo {author} {\bibfnamefont {X.}~\bibnamefont
  {Gao}}, \bibinfo {author} {\bibfnamefont {D.}~\bibnamefont {Paneru}},
  \bibinfo {author} {\bibfnamefont {F.}~\bibnamefont {Di~Colandrea}}, \bibinfo
  {author} {\bibfnamefont {Y.}~\bibnamefont {Zhang}},\ and\ \bibinfo {author}
  {\bibfnamefont {E.}~\bibnamefont {Karimi}},\ }\bibfield  {title} {\bibinfo
  {title} {Generation of the complete bell basis via hong-ou-mandel
  interference of vector modes},\ }\href {https://doi.org/10.1103/dcxb-ql1r}
  {\bibfield  {journal} {\bibinfo  {journal} {Physical Review A}\ }\textbf
  {\bibinfo {volume} {112}},\ \bibinfo {pages} {012215} (\bibinfo {year}
  {2025})}\BibitemShut {NoStop}%
\bibitem [{\citenamefont {Valencia}\ \emph {et~al.}(2026)\citenamefont
  {Valencia}, \citenamefont {Ma}, \citenamefont {Goel}, \citenamefont
  {Leedumrongwatthanakun}, \citenamefont {Graffitti}, \citenamefont {Fedrizzi},
  \citenamefont {McCutcheon},\ and\ \citenamefont {Malik}}]{valencia2026large}%
  \BibitemOpen
  \bibfield  {author} {\bibinfo {author} {\bibfnamefont {N.~H.}\ \bibnamefont
  {Valencia}}, \bibinfo {author} {\bibfnamefont {A.}~\bibnamefont {Ma}},
  \bibinfo {author} {\bibfnamefont {S.}~\bibnamefont {Goel}}, \bibinfo {author}
  {\bibfnamefont {S.}~\bibnamefont {Leedumrongwatthanakun}}, \bibinfo {author}
  {\bibfnamefont {F.}~\bibnamefont {Graffitti}}, \bibinfo {author}
  {\bibfnamefont {A.}~\bibnamefont {Fedrizzi}}, \bibinfo {author}
  {\bibfnamefont {W.}~\bibnamefont {McCutcheon}},\ and\ \bibinfo {author}
  {\bibfnamefont {M.}~\bibnamefont {Malik}},\ }\bibfield  {title} {\bibinfo
  {title} {A large-scale reconfigurable multiplexed quantum photonic network},\
  }\href {https://doi.org/10.1038/s41566-025-01806-x} {\bibfield  {journal}
  {\bibinfo  {journal} {Nature Photonics}\ }\textbf {\bibinfo {volume} {20}},\
  \bibinfo {pages} {202} (\bibinfo {year} {2026})}\BibitemShut {NoStop}%
\bibitem [{\citenamefont {{\.Z}ukowski}\ \emph {et~al.}(1997)\citenamefont
  {{\.Z}ukowski}, \citenamefont {Zeilinger},\ and\ \citenamefont
  {Horne}}]{Zukowski1997}%
  \BibitemOpen
  \bibfield  {author} {\bibinfo {author} {\bibfnamefont {M.}~\bibnamefont
  {{\.Z}ukowski}}, \bibinfo {author} {\bibfnamefont {A.}~\bibnamefont
  {Zeilinger}},\ and\ \bibinfo {author} {\bibfnamefont {M.~A.}\ \bibnamefont
  {Horne}},\ }\bibfield  {title} {\bibinfo {title} {Realizable
  higher-dimensional two-particle entanglements via multiport beam splitters},\
  }\href {https://doi.org/10.1103/PhysRevA.55.2564} {\bibfield  {journal}
  {\bibinfo  {journal} {Physical Review A}\ }\textbf {\bibinfo {volume} {55}},\
  \bibinfo {pages} {2564} (\bibinfo {year} {1997})}\BibitemShut {NoStop}%
\bibitem [{\citenamefont {Ying}\ \emph {et~al.}(2026)\citenamefont {Ying},
  \citenamefont {Zhou}, \citenamefont {Pan}, \citenamefont {Hogan},
  \citenamefont {Zhang}, \citenamefont {Liu}, \citenamefont {Zhu},\ and\
  \citenamefont {Gao}}]{ying2026topological}%
  \BibitemOpen
  \bibfield  {author} {\bibinfo {author} {\bibfnamefont {T.}~\bibnamefont
  {Ying}}, \bibinfo {author} {\bibfnamefont {Y.}~\bibnamefont {Zhou}}, \bibinfo
  {author} {\bibfnamefont {C.}~\bibnamefont {Pan}}, \bibinfo {author}
  {\bibfnamefont {R.}~\bibnamefont {Hogan}}, \bibinfo {author} {\bibfnamefont
  {R.}~\bibnamefont {Zhang}}, \bibinfo {author} {\bibfnamefont
  {H.}~\bibnamefont {Liu}}, \bibinfo {author} {\bibfnamefont {S.}~\bibnamefont
  {Zhu}},\ and\ \bibinfo {author} {\bibfnamefont {X.}~\bibnamefont {Gao}},\
  }\bibfield  {title} {\bibinfo {title} {Topological quantum interferometry},\
  }\bibfield  {journal} {\bibinfo  {journal} {arXiv preprint arXiv:2606.19730}\
  }\href {https://doi.org/10.48550/arXiv.2606.19730}
  {10.48550/arXiv.2606.19730} (\bibinfo {year} {2026})\BibitemShut {NoStop}%
\end{thebibliography}%

%\newpage
\clearpage

%\onecolumn
\onecolumngrid

\setcounter{equation}{0}
\setcounter{figure}{0}
\setcounter{table}{0}
\setcounter{section}{0}

% Redefine numbering format with "S" prefix
\renewcommand{\theequation}{S\arabic{equation}}
\renewcommand{\thefigure}{S\arabic{figure}}
\renewcommand{\thetable}{S\arabic{table}}
\renewcommand{\thesection}{S\arabic{section}}

\section*{Supplementary Material for: Multiphoton Entanglement via Structured-Photon Interference}
\section{q-plate transformation in the local $H/V$ basis}

We first consider a fully tuned q-plate that transforms input photons into vector modes. We give the explicit derivation of the local
$H/V$-basis form, which can be written out using the circular polarization convention 
% \textcolor{red}{S1 and S2 choose one}
\begin{equation}
    \ket{H}
    =
    \frac{1}{\sqrt{2}}
    \left(
    \ket{L}+\ket{R}
    \right),
    \qquad
    \ket{V}
    =
    \frac{i}{\sqrt{2}}
    \left(
    \ket{L}-\ket{R}
    \right).
\label{eq:S_HV_circular}
\end{equation}
% Or equivalently,
% \begin{equation}
%     \ket{R}
%     =
%     \frac{1}{\sqrt{2}}
%     \left(
%     \ket{H}+i\ket{V}
%     \right),
%     \qquad
%     \ket{L}
%     =
%     \frac{1}{\sqrt{2}}
%     \left(
%     \ket{H}-i\ket{V}
%     \right).
% \label{eq:S_circular_HV}
% \end{equation}
% \begin{equation}
%     \ket{L,0}
%     \longrightarrow
%     \ket{R,-2q},
%     \qquad
%     \ket{R,0}
%     \longrightarrow
%     \ket{L,+2q}.
% \label{eq:S_qplate_circular}
% \end{equation}
For a fully tuned ($\delta=\pi$) $q$-plate with topological charge $q$ at an output transverse coordinate $(r,\theta)$, with $H$ or $V$-polarized input photon, modes are written as~\cite{Marrucci2006,Gao2025BellBasis} 
% \begin{equation}
%     \ket{R,-2q}
%     \longrightarrow
%     F_q(r)e^{-i2q\theta}\ket{R},
%     \qquad
%     \ket{L,+2q}
%     \longrightarrow
%     F_q(r)e^{i2q\theta}\ket{L}.
% \label{eq:S_oam_spatial}
% \end{equation}
% For an $H$-polarized input photon, \textcolor{red}{combine S5 and S6}
\begin{align}
    &\ket{H}
    \longrightarrow
    F_q(r)
    \left[
    \cos(2q\theta)\ket{H}
    +
    \sin(2q\theta)\ket{V}
    \right],
    \nonumber\\
    &  \ket{V}
        \longrightarrow
    F_q(r)
    \left[
    \sin(2q\theta)\ket{H}
    -
    \cos(2q\theta)\ket{V}
    \right].
\label{eq:S_H_input_transform}
\end{align}
% Similarly, for a $V$-polarized input photon,
% \begin{align}
%     \ket{V}
%     &=
%     \frac{i}{\sqrt{2}}
%     \left(
%     \ket{L}-\ket{R}
%     \right)
% \longrightarrow
%     \frac{iF_q(r)}{\sqrt{2}}
%     \left(
%     e^{-i2q\theta}\ket{R}
%     -
%     e^{i2q\theta}\ket{L}
%     \right)
%     \nonumber\\
%     &  \ket{V}=
%     F_q(r)
%     \left[
%     \sin(2q\theta)\ket{H}
%     -
%     \cos(2q\theta)\ket{V}
%     \right].
% \label{eq:S_V_input_transform}
% \end{align}
% Equations~\eqref{eq:S_H_input_transform} and
% \eqref{eq:S_V_input_transform} give the local polarization factors used in the
% main text.
For an input photon in port $x$ with q-plate charge $q_x$, detected at
output port $j$, we define
\begin{equation}
    C_{xj}
    =
    \cos(2q_x\theta_j),
    \qquad
    S_{xj}
    =
    \sin(2q_x\theta_j).
\label{eq:S_CS_definition}
\end{equation}
If the input polarization is $p_x\in\{H,V\}$, the local projection factors are therefore
\begin{equation}
    T_{xj}^{(H)}
    =
    \begin{cases}
        C_{xj}, & p_x=H,\\
        S_{xj}, & p_x=V,
    \end{cases}
    \qquad
    T_{xj}^{(V)}
    =
    \begin{cases}
        S_{xj}, & p_x=H,\\
        -C_{xj}, & p_x=V.
    \end{cases}
\label{eq:S_T_factors}
\end{equation}
Note that common radial factors are omitted to better highlight azimuthal dependencies. Although we omit the radial envelope $F_{q_x}(r_j)$, it may be included when absolute spatial intensities are required. Any experimental implementation will also depend on radial mode overlap. 

\section{Bunching probability}
% and the Gram-matrix permanent}

Starting with bunching, let us consider $N$ photons injected into $N$ input ports. The internal state of
the photon entering port $k$ is denoted by $\ket{\chi_k}$, and the
corresponding creation operator is
$\hat a_{k,\chi_k}^{\dagger}$. Hence, the input state is
\begin{equation}
    \ket{\psi_{\mathrm{in}}}
    =
    \prod_{k=1}^{N}
    \hat a_{k,\chi_k}^{\dagger}
    \ket{0}.
\label{eq:S_input_state}
\end{equation}
The interferometric transformation follows
\begin{equation}
    \hat a_{k,\chi_k}^{\dagger}
    \longrightarrow
    \sum_{j=1}^{N}
    U_{kj}
    \hat b_{j,\chi_k}^{\dagger}.
\label{eq:S_interferometer_transform}
\end{equation}
We use standard multiport transformation modelling a 
symmetric Fourier interferometer~\cite{Zukowski1997}, 
\begin{equation}
    U_{kj}
    =
    \frac{1}{\sqrt{N}}
    \omega^{(k-1)(j-1)},
    \qquad
    \omega=e^{2\pi i/N}.
\label{eq:fourier_matrix}
\end{equation}
As such, the output state before postselection is therefore
\begin{equation}
    \ket{\psi_{\mathrm{out}}}
    =
    \prod_{k=1}^{N}
    \left(
    \sum_{j=1}^{N}
    U_{kj}
    \hat b_{j,\chi_k}^{\dagger}
    \right)
    \ket{0}.
\label{eq:S_output_state}
\end{equation}

It is known that for photon bunching to occur, all photons must exit from the same output port. Therefore, the situation that corresponds to this event is in which all photons are found in output port $j$, as follows
\begin{equation}
    \ket{\Phi_j}
    =
    \left(
    \prod_{k=1}^{N}
    U_{kj}
    \right)
    \prod_{k=1}^{N}
    \hat b_{j,\chi_k}^{\dagger}
    \ket{0},
\label{eq:S_bunch_port}
\end{equation}
resulting in the fully bunched component
\begin{equation}
\ket{\widetilde{\psi}_{\mathrm{B}}}
    =
    \sum_{j=1}^{N}
    \left(
    \prod_{k=1}^{N}
    U_{kj}
    \right)
    \prod_{k=1}^{N}
    \hat b_{j,\chi_k}^{\dagger}
    \ket{0}.
\label{eq:S_bunch_state}
\end{equation}
For photons occupying the same output port, the creation and annihilation
operators satisfy
\begin{equation}
    \left[
    \hat b_{j,\chi_k},
    \hat b_{j,\chi_l}^{\dagger}
    \right]
    =
    \braket{\chi_k}{\chi_l}
    =
    G_{kl},
\label{eq:S_commutator_same_port}
\end{equation}
where $G_{kl}$ is the internal-state Gram matrix.  The norm of the internal bunched state is then
\begin{align}
    \bra{0}
    \prod_{k=1}^{N}
    \hat b_{j,\chi_k}
    \prod_{l=1}^{N}
    \hat b_{j,\chi_l}^{\dagger}
    \ket{0}
    =
    \sum_{\pi\in S_N}
    \prod_{k=1}^{N}
    G_{k,\pi(k)}
    =
    \perm G,
\label{eq:S_bunch_norm_permanent}
\end{align}
where $\perm{G}$ represents the permanent of the Gram matrix. Therefore, the probability that all photons exit from output port $j$ is
\begin{equation}
    P_j
    =
    \left|
    \prod_{k=1}^{N}
    U_{kj}
    \right|^2
    \perm G.
\label{eq:S_Pj_bunch}
\end{equation}
Since different bunched output ports correspond to orthogonal photon-number
configurations, the total bunching probability is
\begin{equation}
    P_{\mathrm{B}}
    =
    \sum_{j=1}^{N}
    P_j
    =
    \sum_{j=1}^{N}
    \left|
    \prod_{k=1}^{N}
    U_{kj}
    \right|^2
    \perm G.
\label{eq:S_P_bunch_general}
\end{equation}
For a Fourier-type multiport interferometer,
% of Fourier type, its unitary action can be described as
% \begin{equation}
%     U_{kj}
%     =
%     \frac{1}{\sqrt{N}}
%     \omega^{(k-1)(j-1)},
%     \qquad
%     \omega=e^{2\pi i/N}.
% \label{eq:S_fourier_matrix}
% \end{equation}
we find that $|U_{kj}|^2=1/N$, hence, $P_j=\frac{\perm G}{N^N}$ and thus,
\begin{equation}
    P_{\mathrm{B}}
    =
    \frac{\perm G}{N^{N-1}}.
\label{eq:S_P_bunch_fourier}
\end{equation}
Carefully considering the above, we find two distinct useful cases. In particular, if all internal states are identical ($q_i = q_j$), then
$G_{kl}=1$ for all $k,l$, and $\perm G = N!$, resulting in 
\begin{equation}
    P_{\mathrm{B}}^{q_i=q_j}
    =
    \frac{N!}{N^{N-1}}.
\label{eq:S_P_bunch_identical}
\end{equation}
{Another case is when the internal states are mutually orthogonal ($q_i \neq q_j$). Here, $G=\mathcal{I}$, such that 
$\perm G=1$. As a result, 
\begin{equation}
    P_{\mathrm{B}}^{q_i \neq q_j}
    =
    \frac{1}{N^{N-1}}.
\label{eq:S_P_bunch_orthogonal}
\end{equation}
 Consequently, distinguishable photons can still exhibit bunching, although the probability is reduced by a factor of $N!$ relative to identical internal states. We now consider how the internal-state structure influences the antibunching scenario.}

\section{Antibunching probability}

In this section, we derive the probability of the antibunching event, in which one photon is
detected in each output port. Starting from Eq.~\eqref{eq:S_output_state}, we obtain the antibunching scenario by selecting terms where the $N$ input photons are assigned to the $N$ output ports as a permutation.

Let $\sigma\in S_N$ denote the permutation such that the photon from input
port $\sigma(j)$ exits from output port $j$. The unnormalized antibunching
state is
\begin{equation}
    \ket{\widetilde{\psi}_{\mathrm{GHZ}}}
    =
    \sum_{\sigma\in S_N}
    \left(
    \prod_{j=1}^{N}
    U_{\sigma(j),j}
    \right)
    \prod_{j=1}^{N}
    \hat b_{j,\chi_{\sigma(j)}}^{\dagger}
    \ket{0}.
\label{eq:S_ab_state}
\end{equation}
The antibunching probability can be calculated following, $P_{\mathrm{GHZ}} =
\bra{\widetilde{\Psi}_{\mathrm{GHZ}}}\widetilde{\Psi}_{\mathrm{GHZ}}\rangle$, such that
\begin{align}
    P_{\mathrm{GHZ}}= \langle \widetilde{\psi}_{\mathrm{GHZ}}\ket{\widetilde{\psi}_{\mathrm{GHZ}}} = 
    \sum_{\sigma,\tau\in S_N}
    \left(
    \prod_{j=1}^{N}
    U_{\sigma(j),j}
    U^{*}_{\tau(j),j}
    \right) \times
    \bra{0}
    \prod_{j=1}^{N}
    \hat b_{j,\chi_{\tau(j)}}
    \prod_{j=1}^{N}
    \hat b_{j,\chi_{\sigma(j)}}^{\dagger}
    \ket{0}.
\label{eq:S_P_ab_intermediate}
\end{align}
Since each output port contains one photon, the expectation value reduces to a product of inner products between the corresponding internal states
\begin{equation}
    \bra{0}
    \prod_{j=1}^{N}
    \hat b_{j,\chi_{\tau(j)}}
    \prod_{j=1}^{N}
    \hat b_{j,\chi_{\sigma(j)}}^{\dagger}
    \ket{0}
    =
    \prod_{j=1}^{N}
    \braket{\chi_{\tau(j)}}{\chi_{\sigma(j)}},
\label{eq:S_ab_contraction}
\end{equation}
which conveniently takes the form of the Gram matrix; the equation now becomes
\begin{equation}
    \prod_{j=1}^{N}
    \braket{\chi_{\tau(j)}}{\chi_{\sigma(j)}}
    =
    \prod_{j=1}^{N}
    G_{\tau(j),\sigma(j)}.
\label{eq:S_ab_overlap_product}
\end{equation}
Hence,
\begin{equation}
    P_{\mathrm{GHZ}}
    =
    \sum_{\sigma,\tau\in S_N}
    \prod_{j=1}^{N}
    U_{\sigma(j),j}
    U^{*}_{\tau(j),j}
    G_{\tau(j),\sigma(j)}.
\label{eq:S_P_ab_general}
\end{equation}

Considering similar cases of interest as the bunching scenario, if internal states are mutually orthogonal ($q_i \neq q_j)$, $G=I$, and only terms
with $\sigma=\tau$ survive. As a result,
\begin{equation}
    P_{\mathrm{GHZ}}^{q_i \neq q_j}
    =
    \sum_{\sigma\in S_N}
    \prod_{j=1}^{N}
    |U_{\sigma(j),j}|^2.
\label{eq:S_P_ab_orthogonal_general}
\end{equation}
Considering the multiport interferometer again as Fourier type, $|U_{kj}|^2=1/N$, such that
\begin{equation}
    P_{\mathrm{GHZ}}^{q_i \neq q_j}
    =
    \frac{N!}{N^N}.
\label{eq:S_P_ab_orthogonal_fourier}
\end{equation}
In the contrary case, if all internal states are identical, $G_{kl}=1$, and the probability of antibunching results in
\begin{align}
    P_{\mathrm{GHZ}}^{q_i = q_j}
    =
    \sum_{\sigma,\tau\in S_N}
    \prod_{j=1}^{N}
    U_{\sigma(j),j}
    U^{*}_{\tau(j),j}=
    \left|
    \sum_{\sigma\in S_N}
    \prod_{j=1}^{N}
    U_{\sigma(j),j}
    \right|^2
    =
    |\perm U_N|^2.
\label{eq:S_P_ab_identical}
\end{align}
Here, the compact form for the antibunching probability results in the modulus squared of the permanent of the unitary. However, the probability differs significantly depending on the number of photons. In the following section, we investigate how even versus odd photon number affects the probability. 

\section{Even-$N$ suppression of identical inputs}
 % antibunching in the Fourier interferometer}

Let us consider identical internal states in a multiport Fourier
interferometer, giving the expression of Eq.~\eqref{eq:S_P_ab_identical}. Such probability goes exactly to zero for even photon numbers since $\perm U_N=0$. We now set out to prove this result.

Firstly, we define the unnormalized
Fourier matrix
\begin{equation}
    V_N
    =
    \left[
    \omega^{mn}
    \right]_{m,n=0}^{N-1},
    \qquad
    \omega=e^{2\pi i/N}.
\label{eq:S_VN}
\end{equation}
With a normalization factor of $N^{-N/2}$, the fully normalized Fourier is $U_N=N^{-1/2}V_N$, such that
\begin{equation}
    \perm U_N
    =
    N^{-N/2}
    \perm V_N.
\label{eq:S_perm_normalization}
\end{equation}
Let $V_N'$ be a cyclical shift of all rows upward by
one row of $V_N$. Since a row permutation does not change the permanent,
\begin{equation}
    \perm V_N'
    =
    \perm V_N,
\label{eq:S_row_perm_invariance}
\end{equation}
but a cyclic row shift acts as
\begin{equation}
    V'_{mn}
    =
    \omega^n V_{mn}.
\label{eq:S_row_shift_phase}
\end{equation}
Therefore, each column $n$ acquires a factor $\omega^n$, and the permanent
changes by
\begin{align}
    \perm V_N'
    &=
    \left(
    \prod_{n=0}^{N-1}
    \omega^n
    \right)
    \perm V_N
    \nonumber\\
    &=
    \omega^{N(N-1)/2}
    \perm V_N
    \nonumber\\
    &=
    e^{\pi i (N-1)}
    \perm V_N
    \nonumber\\
    &=
    (-1)^{N-1}
    \perm V_N.
\label{eq:S_perm_phase}
\end{align}
Combining Eqs.~\eqref{eq:S_row_perm_invariance} and
\eqref{eq:S_perm_phase}, we obtain
\begin{equation}
    \perm V_N
    =
    (-1)^{N-1}
    \perm V_N.
\end{equation}
For even $N$, $(-1)^{N-1}=-1$, so
\begin{equation}
    \perm V_N=0,
    \qquad
    \perm U_N=0.
\label{eq:S_even_N_suppression}
\end{equation}
Thus, for identical photons entering an even-dimensional Fourier
interferometer, the antibunching probability vanishes
\begin{equation}
    P_{\mathrm{GHZ}}
    =
    |\perm U_N|^2
    =
    0,
    \qquad
    N \ \mathrm{even}.
\label{eq:S_even_N_Pab_zero}
\end{equation}
In the following sections, we examine the overall general polarization structure and a three-photon case study.

\section{Polarization-string amplitudes and GHZ-like decomposition}

Consider the local-polarization $H/V$-basis expansion of the antibunching output state, from which we can derive corresponding GHZ-like decomposition. We start by examining each output port $j$, such that the internal state of the photon originating from input
port $k$ can be written as
\begin{equation}
    \ket{\chi_k}_j
    \longrightarrow
    X_{kj}^{(H)}
    \ket{H}_j
    +
    X_{kj}^{(V)}
    \ket{V}_j.
\label{eq:S_local_HV_expansion}
\end{equation}
For an $H$-polarized input photon, the coefficients are
\begin{equation}
    X_{kj}^{(H)}
    =
    F_{q_k}(r_j)
    \cos(2q_k\theta_j),
    \qquad
    X_{kj}^{(V)}
    =
    F_{q_k}(r_j)
    \sin(2q_k\theta_j).
\label{eq:S_X_coefficients_H_input}
\end{equation}
For a general input polarization, the coefficients are given by the local
factors in Eq.~\eqref{eq:S_T_factors}.
Here, we define a polarization string as
\begin{equation}
    \bm{\pi}
    =
    (\pi_1,\pi_2,\ldots,\pi_N),
    \qquad
    \pi_j\in\{H,V\},
\label{eq:S_pol_string}
\end{equation}
where $\pi_j$ is the polarization at output port $j$. The corresponding basis state is therefore
\begin{equation}
    \ket{\bm{\pi}}
    =
    \ket{\pi_1}_1
    \ket{\pi_2}_2
    \cdots
    \ket{\pi_N}_N.
\label{eq:S_pol_basis}
\end{equation}
Substituting Eq.~\eqref{eq:S_local_HV_expansion} into Eq.~\eqref{eq:S_ab_state}, the amplitude of the polarization string
$\ket{\bm{\pi}}$ is
\begin{equation}
    A_{\bm{\pi}}
    =
    \sum_{\sigma\in S_N}
    \prod_{j=1}^{N}
    U_{\sigma(j),j}
    X_{\sigma(j),j}^{(\pi_j)}.
\label{eq:S_pol_string_amplitude}
\end{equation}
The unnormalized antibunching state can therefore be written as
\begin{equation}
    \ket{\widetilde{\psi}_{\mathrm{GHZ}}}
    =
    \sum_{\bm{\pi}\in\{H,V\}^{N}}
    A_{\bm{\pi}}
    \ket{\bm{\pi}}.
\label{eq:S_HV_expansion}
\end{equation}
For any polarization string defined by Eq.~\eqref{eq:S_pol_string}, we have a complementary string
$\overline{\bm{\pi}}$, which is obtained by exchanging polarizations $H\leftrightarrow V$
at every output port. In this form, we can now define corresponding GHZ-like states as
\begin{equation}
    \ket{\mathrm{GHZ}_{\bm{\pi}}^{\pm}}
    =
    \frac{1}{\sqrt{2}}
    \left(
    \ket{\bm{\pi}}
    \pm
    \ket{\overline{\bm{\pi}}}
    \right).
\label{eq:S_GHZ_like_basis}
\end{equation}
For each complementary pair, the corresponding coefficients are $s_{\bm{\pi}}^{\pm}=\frac{1}{\sqrt{2}}\left(A_{\bm{\pi}}\pm A_{\overline{\bm{\pi}}}\right)$, which, when squared, give raw component weights as
\begin{equation}
    P_{\bm{\pi}}^{\pm}
    =
    |s_{\bm{\pi}}^{\pm}|^2
    =
    \frac{1}{2}
    \left|
    A_{\bm{\pi}}
    \pm
    A_{\overline{\bm{\pi}}}
    \right|^2.
\label{eq:S_GHZ_weights}
\end{equation}
Such raw weights,
$P_{\bm{\pi}}^{\pm}$, are the primary focus of the main text, as they directly display the spatial distribution of
each GHZ-like component. However, normalized GHZ-like probabilities are also obtained by dividing by the total antibunching probability, as $p_{\bm{\pi}}^{\pm}=\frac{P_{\bm{\pi}}^{\pm}}{P_{\mathrm{GHZ}}}$.

\section{Three-photon analysis}
\label{sec:S_three_photon_formulas}

We now consider a three-photon case study. We begin with the antibunching case and give a detailed analysis surrounding the explicit three-photon polarization string amplitudes. We first take input (output) ports labeled by $a,b,c$ ($d,e,f$), and use the tritter convention
\begin{equation}
    U^{(3)}
    =
    \frac{1}{\sqrt{3}}
    \begin{pmatrix}
        1 & 1 & 1 \\
        1 & \zeta^2 & \zeta \\
        1 & \zeta & \zeta^2
    \end{pmatrix},
    \qquad
    \zeta=e^{2\pi i/3}.
    \label{eq:S_tritter_convention}
\end{equation}
corresponding to rows (columns) as input (output) ports. The product of three tritter amplitudes gives the common factor $1/(3\sqrt{3})$. Using this matrix, we define an unnormalized three-photon antibunching state as
\begin{equation}
    |\widetilde{\psi}_{111}\rangle
    =
    \sum_{\sigma\in S_3}
    \left(
    \prod_{j\in\{d,e,f\}}
    U_{\sigma(j),j}
    \right)
    \prod_{j\in\{d,e,f\}}
    \hat b_{j,\chi_{\sigma(j)}}^{\dagger}
    |0\rangle .
    \label{eq:S_three_ab_state}
\end{equation}
Expanding it in the local $H/V$ polarization basis gives
\begin{equation}
    |\widetilde{\psi}_{111}\rangle
    =
    \sum_{\boldsymbol{\pi}\in\{H,V\}^{3}}
    A_{\boldsymbol{\pi}}
    |\pi_d\pi_e\pi_f\rangle ,
    \label{eq:S_three_HV_expansion}
\end{equation}
where $\boldsymbol{\pi}=(\pi_d,\pi_e,\pi_f)$ is the polarization string associated with the output ports $d,e,f$. Input photons in ports $x$, $x\in\{a,b,c\}$,  passing through q-plates with charge $q_x$ and detected at output port $j\in\{d,e,f\}$ follow the trignometric functions as defined in Eqs.~\eqref{eq:S_CS_definition} and ~\eqref{eq:S_T_factors}.
%  we define the local trigonometric
% factors
% \begin{equation}
%     C_{xj}
%     =
%     \cos(2q_x\theta_j),
%     \qquad
%     S_{xj}
%     =
%     \sin(2q_x\theta_j).
%     \label{eq:S_three_CS}
% \end{equation}
% If the input polarization in port $x$ is $p_x=H$, the local projection
% factors are
% \begin{equation}
%     X_{xj}^{H}=C_{xj},
%     \qquad
%     X_{xj}^{V}=S_{xj}.
%     \label{eq:S_three_X_H_input}
% \end{equation}
% If the input polarization in port $x$ is $p_x=V$, they become
% \begin{equation}
%     X_{xj}^{H}=S_{xj},
%     \qquad
%     X_{xj}^{V}=-C_{xj}.
%     \label{eq:S_three_X_V_input}
% \end{equation}
% The radial envelopes $F_{q_x}(r_j)$ can be included in
% $X_{xj}^{H,V}$ when absolute spatial intensities are required. In the angular
% plots of the main text, we focus on the azimuthal dependence and omit common
% radial factors.
For a general polarization string, $\boldsymbol{\pi}$, the amplitude is obtained
by summing the six indistinguishable three-photon paths
\begin{align}
    A_{\pi_d\pi_e\pi_f}
    =
    \frac{1}{3\sqrt{3}}
    \Big(
    &\zeta
    X_{ad}^{\pi_d}
    X_{be}^{\pi_e}
    X_{cf}^{\pi_f}
    +
    \zeta^2
    X_{ad}^{\pi_d}
    X_{ce}^{\pi_e}
    X_{bf}^{\pi_f}+
    \zeta^2
    X_{bd}^{\pi_d}
    X_{ae}^{\pi_e}
    X_{cf}^{\pi_f}
    +
    \zeta
    X_{bd}^{\pi_d}
    X_{ce}^{\pi_e}
    X_{af}^{\pi_f}
    +
    \zeta
    X_{cd}^{\pi_d}
    X_{ae}^{\pi_e}
    X_{bf}^{\pi_f}
    +
    \zeta^2
    X_{cd}^{\pi_d}
    X_{be}^{\pi_e}
    X_{af}^{\pi_f}
    \Big),
\label{eq:S_three_general_string_amplitude}
\end{align}
where six input-output assignments correspond to $
(a,b,c),(a,c,b), (b,a,c),
(b,c,a),(c,a,b),(c,b,a)
$
over the output ports $(d,e,f)$, respectively. The eight polarization-string amplitudes are therefore
\begin{align}
    A_{HHH}
    =
    \frac{1}{3\sqrt{3}}
    \Big(
    &\zeta X_{ad}^{H}X_{be}^{H}X_{cf}^{H}
    +
    \zeta^2 X_{ad}^{H}X_{ce}^{H}X_{bf}^{H}
    +
    \zeta^2 X_{bd}^{H}X_{ae}^{H}X_{cf}^{H}+
    \zeta X_{bd}^{H}X_{ce}^{H}X_{af}^{H}
    +
    \zeta X_{cd}^{H}X_{ae}^{H}X_{bf}^{H}
    +
    \zeta^2 X_{cd}^{H}X_{be}^{H}X_{af}^{H}
    \Big),
    \label{eq:S_A_HHH}
    \\
    A_{HHV}
    =
    \frac{1}{3\sqrt{3}}
    \Big(
    &\zeta X_{ad}^{H}X_{be}^{H}X_{cf}^{V}
    +
    \zeta^2 X_{ad}^{H}X_{ce}^{H}X_{bf}^{V}
    +
    \zeta^2 X_{bd}^{H}X_{ae}^{H}X_{cf}^{V}+
    \zeta X_{bd}^{H}X_{ce}^{H}X_{af}^{V}
    +
    \zeta X_{cd}^{H}X_{ae}^{H}X_{bf}^{V}
    +
    \zeta^2 X_{cd}^{H}X_{be}^{H}X_{af}^{V}
    \Big),
    \label{eq:S_A_HHV}
    \\
    A_{HVH}
    =
    \frac{1}{3\sqrt{3}}
    \Big(
    &\zeta X_{ad}^{H}X_{be}^{V}X_{cf}^{H}
    +
    \zeta^2 X_{ad}^{H}X_{ce}^{V}X_{bf}^{H}
    +
    \zeta^2 X_{bd}^{H}X_{ae}^{V}X_{cf}^{H}+
    \zeta X_{bd}^{H}X_{ce}^{V}X_{af}^{H}
    +
    \zeta X_{cd}^{H}X_{ae}^{V}X_{bf}^{H}
    +
    \zeta^2 X_{cd}^{H}X_{be}^{V}X_{af}^{H}
    \Big),
    \label{eq:S_A_HVH}
    \\
    A_{HVV}
    =
    \frac{1}{3\sqrt{3}}
    \Big(
    &\zeta X_{ad}^{H}X_{be}^{V}X_{cf}^{V}
    +
    \zeta^2 X_{ad}^{H}X_{ce}^{V}X_{bf}^{V}
    +
    \zeta^2 X_{bd}^{H}X_{ae}^{V}X_{cf}^{V}+
    \zeta X_{bd}^{H}X_{ce}^{V}X_{af}^{V}
    +
    \zeta X_{cd}^{H}X_{ae}^{V}X_{bf}^{V}
    +
    \zeta^2 X_{cd}^{H}X_{be}^{V}X_{af}^{V}
    \Big),
    \label{eq:S_A_HVV}
    \\
    A_{VHH}
    =
    \frac{1}{3\sqrt{3}}
    \Big(
    &\zeta X_{ad}^{V}X_{be}^{H}X_{cf}^{H}
    +
    \zeta^2 X_{ad}^{V}X_{ce}^{H}X_{bf}^{H}
    +
    \zeta^2 X_{bd}^{V}X_{ae}^{H}X_{cf}^{H}+
    \zeta X_{bd}^{V}X_{ce}^{H}X_{af}^{H}
    +
    \zeta X_{cd}^{V}X_{ae}^{H}X_{bf}^{H}
    +
    \zeta^2 X_{cd}^{V}X_{be}^{H}X_{af}^{H}
    \Big),
    \label{eq:S_A_VHH}
    \\
    A_{VHV}
    =
    \frac{1}{3\sqrt{3}}
    \Big(
    &\zeta X_{ad}^{V}X_{be}^{H}X_{cf}^{V}
    +
    \zeta^2 X_{ad}^{V}X_{ce}^{H}X_{bf}^{V}
    +
    \zeta^2 X_{bd}^{V}X_{ae}^{H}X_{cf}^{V}+
    \zeta X_{bd}^{V}X_{ce}^{H}X_{af}^{V}
    +
    \zeta X_{cd}^{V}X_{ae}^{H}X_{bf}^{V}
    +
    \zeta^2 X_{cd}^{V}X_{be}^{H}X_{af}^{V}
    \Big),
    \label{eq:S_A_VHV}
    \\
    A_{VVH}
    =
    \frac{1}{3\sqrt{3}}
    \Big(
    &\zeta X_{ad}^{V}X_{be}^{V}X_{cf}^{H}
    +
    \zeta^2 X_{ad}^{V}X_{ce}^{V}X_{bf}^{H}
    +
    \zeta^2 X_{bd}^{V}X_{ae}^{V}X_{cf}^{H}+
    \zeta X_{bd}^{V}X_{ce}^{V}X_{af}^{H}
    +
    \zeta X_{cd}^{V}X_{ae}^{V}X_{bf}^{H}
    +
    \zeta^2 X_{cd}^{V}X_{be}^{V}X_{af}^{H}
    \Big),
    \label{eq:S_A_VVH}
    \\
    A_{VVV}
    =
    \frac{1}{3\sqrt{3}}
    \Big(
    &\zeta X_{ad}^{V}X_{be}^{V}X_{cf}^{V}
    +
    \zeta^2 X_{ad}^{V}X_{ce}^{V}X_{bf}^{V}
    +
    \zeta^2 X_{bd}^{V}X_{ae}^{V}X_{cf}^{V}+
    \zeta X_{bd}^{V}X_{ce}^{V}X_{af}^{V}
    +
    \zeta X_{cd}^{V}X_{ae}^{V}X_{bf}^{V}
    +
    \zeta^2 X_{cd}^{V}X_{be}^{V}X_{af}^{V}
    \Big).
    \label{eq:S_A_VVV}
\end{align}
The three-photon GHZ-like basis is obtained by pairing complementary
polarization strings
\begin{align}
    |S_1^\pm\rangle
    &=
    \frac{1}{\sqrt{2}}
    \left(
    |HHH\rangle
    \pm
    |VVV\rangle
    \right),
    \qquad
    |S_2^\pm\rangle
    =
    \frac{1}{\sqrt{2}}
    \left(
    |HHV\rangle
    \pm
    |VVH\rangle
    \right),
    \nonumber\\
    |S_3^\pm\rangle
    &=
    \frac{1}{\sqrt{2}}
    \left(
    |HVH\rangle
    \pm
    |VHV\rangle
    \right),
    \qquad
    |S_4^\pm\rangle
    =
    \frac{1}{\sqrt{2}}
    \left(
    |HVV\rangle
    \pm
    |VHH\rangle
    \right),
    \label{eq:S_three_GHZ_basis}
\end{align}
with corresponding GHZ-like coefficients as
\begin{align}
    S_1^\pm
    &=
    \frac{1}{\sqrt{2}}
    \left(
    A_{HHH}
    \pm
    A_{VVV}
    \right),
    \qquad
    S_2^\pm=
    \frac{1}{\sqrt{2}}
    \left(
    A_{HHV}
    \pm
    A_{VVH}
    \right),
    \nonumber\\
    S_3^\pm
    &=
    \frac{1}{\sqrt{2}}
    \left(
    A_{HVH}
    \pm
    A_{VHV}
    \right),
    \qquad
    S_4^\pm
    =
    \frac{1}{\sqrt{2}}
    \left(
    A_{HVV}
    \pm
    A_{VHH}
    \right),
    \label{eq:S_three_GHZ_coefficients}
\end{align}
such that raw GHZ-like component weights are $P_{S_\mu^\pm}=|S_\mu^\pm|^2$ with $\mu=1,2,3,4$.

%\bibliography{refs}

\end{document}